\documentclass[conference,compsoc,letterpaper]{IEEEtran}

\usepackage{tikz}
\usetikzlibrary{positioning, arrows.meta, fit, backgrounds, calc}
\usepackage{amsmath}
\usepackage{booktabs}
\usepackage{multirow}
\usepackage{hhline}
\usepackage[table]{xcolor}
\usepackage{listings}
\usepackage{graphicx}
\usepackage{xurl}
\usepackage{balance}
\usepackage{comment}
\usepackage{circledsteps}
\usepackage{xspace}
\usepackage{pifont}
\usepackage{orcidlink}
\usepackage{enumitem}
\usepackage[most]{tcolorbox}
\tcbuselibrary{listings}
\usepackage{hyperref}
\usepackage{cite}
\usepackage{amssymb} 
\usepackage{array}
\usepackage{longtable}
\usepackage{empheq}
\usepackage[x11names]{xcolor} % For custom colors
\hypersetup{hidelinks}
\usepackage{enumitem}

\newlist{researchquestions}{enumerate}{1}
\setlist[researchquestions]{
    label=\textbf{RQ\arabic*:},
    ref=RQ\arabic*,
    leftmargin=*,
    itemsep=0.5em
}

\ifCLASSOPTIONcompsoc
 \usepackage[caption=false,font=normalsize,labelfont=sf,textfont=sf]{subfig}
\else
 \usepackage[caption=false,font=footnotesize]{subfig}
\fi

\newcommand{\stepnum}[1]{%
\tikz[baseline=(char.base)]{
\node[
    circle,
    fill=black,
    text=white,
    inner sep=2.0pt,
    minimum size=3mm,
    font=\bfseries\small
] (char) {#1};}}

\definecolor{fieldbg}{HTML}{D6E4F0}

\definecolor{selectionbg}{HTML}{F7E9B5}
\newcommand{\selection}[1]{%
  \colorbox{selectionbg}{\strut#1}%
}

\definecolor{capabilitybg}{HTML}{D7CEE3}
\newcommand{\capability}[1]{%
  \colorbox{capabilitybg}{\strut#1}%
}

\definecolor{detectionbg}{HTML}{D5DFF3}
\newcommand{\detection}[1]{%
  \colorbox{detectionbg}{\strut#1}%
}

\newtcblisting{minimalprompt}{
    listing only,
    breakable,                % Allows the box to break across pages if long
    blanker,                  % Removes default borders and margins
    left=12pt,                % Left padding for the vertical rule
    before skip=12pt,         
    after skip=12pt,          
    borderline left={2pt}{0pt}{gray!60}, % Clean gray vertical accent line
    colback=gray!4,           % Subtle background fill
    width=\textwidth,         % Explicitly force full text width
    listing options={
        basicstyle=\small\ttfamily\color{black!90},
        breaklines=true,       
        columns=fullflexible
    }
}

\newcommand{\tier}[1]{\textsc{Tier\,#1}}
\newcommand{\benign}{\textsc{benign}}
\newcommand{\suspicious}{\textsc{suspicious}}
\newcommand{\malicious}{\textsc{malicious}}

\newcommand{\efren}[1]{\textcolor{magenta}{\bf {\bf Efren:}  #1}}

\newcommand{\framework}{\textsc{SecCARD}\xspace}

\newcolumntype{L}{>{\raggedright\arraybackslash}p{4.2cm}}
\newcolumntype{R}{>{\raggedleft\arraybackslash}p{2.5cm}}
\renewcommand{\arraystretch}{1.15}

\definecolor{attackred}{RGB}{220,50,47}

\begin{document}
\title{We Have All the Cards: Securing Agent Cards \\at the  Operational Trust Boundary of the Agent2Agent Protocol}

% author names and affiliations use a multiple column layout for up to three different affiliations

% \author{\IEEEauthorblockN{Michael Shell}
% \IEEEauthorblockA{School of Electrical and\\Computer Engineering\\
% Georgia Institute of Technology\\
% Atlanta, Georgia 30332--0250\\
% Email: http://www.michaelshell.org/contact.html}
% \and
% \IEEEauthorblockN{Homer Simpson}
% \IEEEauthorblockA{Twentieth Century Fox\\
% Email: homer@thesimpsons.com}
% \and
% \IEEEauthorblockN{James Kirk\\ and Montgomery Scott}
% \IEEEauthorblockA{Starfleet Academy\\
% San Francisco, California 96678--2391\\
% Telephone: (800) 555--1212\\

%Telephone: (800) 555--1212, Fax: (888) 555--1212}

\author{
\IEEEauthorblockN{Miguel A. Guirao Aguilera}
\IEEEauthorblockA{New Mexico State University\\
guirao@nmsu.edu}
\and
\IEEEauthorblockN{Satyajayant Misra}
\IEEEauthorblockA{The University of Texas at Dallas \\
jay.misra@utdallas.edu}
\and
\IEEEauthorblockN{Efrén López-Morales }
\IEEEauthorblockA{New Mexico State University\\
elopezm@nmsu.edu}
}

\maketitle

\begin{abstract}
AI agent interoperability protocols enable orchestrators to discover
and invoke external agents using machine-readable metadata. In the
Agent2Agent (A2A) protocol, Agent Cards contain attacker-controllable
natural-language fields that may influence agent selection before a
provider is invoked, creating a pre-invocation trust problem. We
characterize this discovery-time attack surface through a taxonomy of
ten Agent Card manipulation attacks and present \framework, a layered
validation framework combining protocol checks, deterministic analysis,
and LLM-based semantic reasoning.

We measure the emerging public A2A ecosystem by probing 36.1 million
Internet domains and collecting 310 deployed Agent Cards. Two
annotators independently label the corpus with 96.8\% agreement. We then construct a 660-card real-card perturbation benchmark with matched benign controls and a 260-card synthetic
adversarial benchmark spanning all ten attack categories. On controlled
attacks injected into real Agent Cards, the strongest semantic judge
achieves 97.0\% security recall at a 1.7\% false-positive rate. Across
nine LLM judges on the synthetic benchmark, security recall ranges
from 25.8\% to 92.3\%, demonstrating substantial model-dependent
variation.

These results show that \framework effectively identifies Agent Card manipulations spanning all ten discovery-time attack classes while preserving benign Agent Cards,
but also reveal that its effectiveness depends critically on the
choice of the underlying LLM judge.
\end{abstract}


% \fi
\IEEEpeerreviewmaketitle

% ================ MAIN BODY STARTS HERE ==================================

% ------------------------------------------------
% INTRODUCTION (OK)
% ------------------------------------------------
% ------------------------------------------------
% INTRODUCTION
% -----------------------------------------------
\section{Introduction}
\label{sec:intro}

Advances in Large Language Models (LLMs) have accelerated the development of autonomous LLM-based AI agents capable of planning and multi-step reasoning. As these systems grow in complexity, organizations are adopting orchestrator agents that coordinate multiple specialized agents to collaboratively solve complex tasks~\cite{chopra2026orchestration,dang2026multi}.

This architectural shift is rapidly moving into production. In a 2026 survey, 57\% of professionals reported having AI agents already deployed in production~\cite{langchain2026state}.

This has motivated the development of standardized interoperability protocols that enable agent discovery, capability advertisement, and communication across agents. Among these, the Agent2Agent (A2A) protocol, released in April 2025 by Google, has emerged as one of the first open standards designed for agent interoperability~\cite{google2025a2a}.

A fundamental challenge faced by agent interoperability protocols is \emph{agent discovery}: before an orchestrator can delegate a task, it must discover candidate agents and determine which one is most suitable, i.e. possesses the required skill(s) to complete the task. Unlike traditional service-oriented architectures, where endpoints are configured manually, agent orchestrators require machine-readable metadata that describe an agent's identity, capabilities, skills, authentication requirements, and communication endpoints.

To solve this challenge, the A2A protocol allows organizations to publish JavaScript Object Notation (JSON)-based \emph{agent cards} to advertise their capabilities to other agents~\cite{a2a2026spec,nottingham2019rfc8615}. An orchestrating agent that wishes to delegate a task fetches the publicly available card file, parses its fields and decides whether to invoke the advertised service, \emph{all without human review}.

The security implication is immediate: \emph{agent card content is untrusted text that an LLM consumes as input.} An attacker who controls a domain or compromises a legitimate one can publish a card whose text fields carry prompt injection payloads targeted at the ingesting orchestrator.
Unlike web content that humans evaluate before acting on, agent cards are designed to drive autonomous LLM-based agent decisions that may trigger real-world actions: transferring funds, exfiltrating conversation context, or silently escalating invocation privileges. The A2A specification provides no normative guidance on how to evaluate card trustworthiness beyond schema conformance.

\noindent\textbf{Research gap.}
Indirect prompt injection through externally controlled content is a
well-established threat to LLM-integrated systems
~\cite{debenedetti2024agentdojo,greshake2023not,liu2024formalizing,zhan2024injecagent}.
Recent work has extended this concern to A2A systems, identifying
Agent Card spoofing, capability cloaking, and card poisoning as
potential attacks on agent discovery.

However, existing work characterizes only a limited set of discovery-time Agent Card manipulations and primarily evaluates attacks in controlled settings. We still lack
a fine-grained understanding of the ways adversarial Agent Cards can
manipulate discovery decisions, empirical evidence about such
manipulations in deployed Agent Cards, and a dedicated
pre-invocation validation approach evaluated against both adversarial
and benign cards.

This gap is operationally important because Agent Cards are not merely
external content retrieved by an LLM; they are metadata explicitly
designed to influence which external agents an orchestrator selects
and invokes. Consequently, adversarial content can affect the
discovery decision before the advertised provider enters the
workflow.

\noindent\textbf{Our approach.}
We study Agent Card security at the discovery-to-invocation trust
boundary, where attacker-controlled metadata can influence an
orchestrator before the advertised provider is invoked. We develop a
taxonomy of ten discovery-time Agent Card manipulation attacks,
characterizing how adversarial cards can manipulate agent selection,
misrepresent capabilities, and evade validation.

Building on this taxonomy, we introduce \framework, a layered
pre-invocation validation framework. Layer~1 validates protocol
conformance against the A2A specification. Layer~2A applies
deterministic NLP analysis to identify lexical and structural
indicators of manipulation, while Layer~2B employs an LLM-as-a-judge
semantic classifier~\cite{zheng2023judging} to identify attacks that
require reasoning about meaning, intent, and consistency.

We evaluate \framework using three complementary datasets: 310
real-world Agent Cards collected through an Internet-scale measurement,
a 660-card real-card perturbation benchmark with matched benign
controls, and a 260-card synthetic adversarial benchmark spanning all
ten attack categories. On controlled attacks injected into real Agent
Cards, the strongest semantic judge achieves 97.0\% security recall
at a 1.7\% false-positive rate. Across nine LLM judges on the
synthetic benchmark, security recall ranges from 25.8\% to 92.3\%.
Together, these experiments evaluate Agent Card validation across
naturally occurring deployments, controlled attacks on real cards,
and broader adversarial variation.

\noindent\textbf{Contributions.}
This paper makes the following contributions:

\begin{itemize}[leftmargin=*,noitemsep]

%\item \textbf{Operational characterization of the Agent Card trust boundary.}
%We characterize Agent Card discovery as a pre-invocation trust problem, where attacker-controlled metadata can influence which external agents are admitted before those agents are invoked.

\item \textbf{Internet-scale measurement of deployed Agent Cards.}
We probe 36.1 million Internet domains and collect 310 publicly deployed
Agent Cards, providing an empirical view of the emerging A2A ecosystem.
Two annotators independently establish security ground truth for the
corpus, achieving 96.8\% agreement ($\kappa=0.804$).

\item \textbf{Agent card manipulation attack taxonomy.}
Drawing on our Internet measurement and prior characterizations of
prompt injection in agentic systems, we develop a taxonomy of ten Agent Card manipulation attacks across three categories: manipulating agent selection,
misrepresenting capabilities, and evading validation.

\item \textbf{\framework: Pre-invocation Agent Card validation.}
Based on our attack taxonomy, we introduce \framework, a layered validation framework
combining protocol conformance checks, deterministic analysis, and
LLM-based semantic reasoning. On controlled attacks injected into
real Agent Cards, the strongest semantic judge achieves 97.0\%
security recall at a 1.7\% false-positive rate.

\item \textbf{Controlled adversarial evaluation.}
We construct a 660-card real-card perturbation benchmark with matched
benign controls and a 260-card synthetic adversarial benchmark spanning
all ten attack categories. Across nine LLM judges, we evaluate detection
effectiveness, false-positive behavior, and model-dependent robustness
under controlled agent card manipulation attacks
\footnote{\url{https://github.com/cps-sec/seccard}}.
\end{itemize}

\section{Related Work}
\label{sec:related}

\textbf{A2A and Agent Card Security.}
Security research on the A2A protocol has begun to characterize
threats across agent discovery, orchestration, and execution.
Habler et al.~\cite{habler2025building} provide a broad threat analysis
of A2A, including risks associated with Agent Card discovery, while
Louck et al.~\cite{louck} focus on privacy and secure
information exchange. He et al.~\cite{he2024survey} and Sharma
et al.~\cite{sharma2025towards} study adversarial behavior and
security evaluation in broader multi-agent systems rather than A2A
discovery specifically.

Most closely related, Li et al.'s
A2ASecBench~\cite{li2026a2asecbench} introduces a security benchmark
spanning the A2A lifecycle. Among its six attack classes, two target
agent admission through Agent Cards: AgentCard Spoofing and Capability
Cloaking. A2ASecBench evaluates whether A2A systems can withstand these
and other attacks, but does not introduce a dedicated pre-invocation
Agent Card validation mechanism. In contrast, our work focuses
specifically on the Agent Card discovery boundary: we develop a
fine-grained taxonomy of ten Agent Card manipulation attacks, measure
deployed Agent Cards in the wild, establish human-annotated security
ground truth, and introduce \framework for pre-invocation Agent Card
validation. Table~\ref{tab:related_work} summarizes these distinctions.

Deshmukh et al.~\cite{deshmukh2026agent} empirically study Agent Card
poisoning, demonstrating that manipulated cards can redirect
orchestrators to attacker-controlled endpoints and evaluating detection
strategies in a controlled testbed. Our work considers a broader set of
discovery-time manipulations and complements controlled adversarial
evaluation with Internet-scale measurement of deployed Agent Cards and
human-established ground truth.

\textbf{Prompt Injection in LLM Applications and Agents.}
Prompt injection allows adversarial instructions embedded in
attacker-controlled text to influence an LLM's behavior
~\cite{perez2022ignore}. Greshake et al.~\cite{greshake2023not}
extended this threat to \emph{indirect} prompt injection, where
malicious instructions originate from externally retrieved content.
Subsequent work has demonstrated related risks in tool-integrated
agents and external resources~\cite{schick2023toolformer,yao2023react}.

Our work studies a distinct point in the agent lifecycle: untrusted
content embedded in \emph{agent discovery metadata}. A2A Agent Cards
are retrieved and interpreted before task execution and can influence
which external agent an orchestrator selects and invokes. We therefore
focus on detecting manipulation at discovery time, before the
advertised provider agent enters the autonomous workflow.

\begin{table*}[t]
\centering
\caption{Comparison of related work on A2A and agent security.}
\label{tab:related_work}
\resizebox{\textwidth}{!}{
\begin{tabular}{lcccccc}
\toprule
\textbf{Work} &
\textbf{A2A} &
\textbf{Agent Card Security} &
\textbf{Agent Card Attack Classes} &
\textbf{Wild Agent Cards} &
\textbf{Human Ground Truth} &
\textbf{Pre-Invocation Defense} \\
\midrule

Habler et al.~\cite{habler2025building}
& \checkmark & \checkmark & -- & -- & -- & -- \\

Louck et al.~\cite{louck}
& \checkmark & -- & -- & -- & -- & -- \\

Li et al.~\cite{li2026a2asecbench}
& \checkmark & \checkmark & 2 & -- & -- & -- \\

\midrule
\textbf{This work}
& \checkmark & \checkmark & \textbf{10} &
\checkmark & \checkmark & \checkmark \\

\bottomrule
\end{tabular}
}
\end{table*}

\section{Background}
\label{sec:background}

This section lays out key relevant concepts including the A2A protocol (Sec.~\ref{sec:problem:a2a}), and the discovery-to-invocation process (Sec.~\ref{subsec:discovery}).

\subsection{The A2A Protocol}
\label{sec:problem:a2a}

The A2A protocol is an open standard for AI agent interoperability~\cite{google2025a2a,a2a2026spec}. A2A is one instance of a broader class of \emph{agent metadata protocols} that expose machine-readable capability descriptions at a predictable Uniform Resource Identifier (URI). Anthropic's Model Context Protocol (MCP)~\cite{mcp2024spec} exposes tool schemas similarly; the emerging \texttt{llms.txt} convention publishes natural-language summaries of site content for LLM consumption.

A2A's central mechanism is a machine-readable \emph{agent card}: a JSON document that an agent server publishes at the URI \texttt{\url{/.well-known/agent-card.json}}, following the RFC~8615 well-known resource convention~\cite{nottingham2019rfc8615}.
A client wishing to delegate work discovers candidate agents via this path, evaluates their declared capabilities, and initiates tasks over  JSON-RPC or REST using the endpoint in the card.

\begin{comment}
    
\textbf{A2A Deployment.} 
For a \textbf{provider agent}  to offer its services to a \textbf{client agent} (also known as orchestrator agent), it must publish its agent card at the well-known location for agents to discover and fetch. Client agents fetch cards at discovery time, parse their contents, and use the provided endpoints to route subsequent calls for services. 
The A2A protocol fosters interoperability by providing a common language for agents built using diverse frameworks such as CrewAI~\cite{crewai_github_2026} or LangChain~\cite{langchain2026}.

\end{comment}
%A2A cards can be consumed by an orchestrator agent developed using any framework such as CrewAI\footnote{https://crewai.com/} or LangChain\footnote{https://www.langchain.com/}.  The A2A protocol is a communication protocol between agents that requires a client or orchestrator agent consuming the services offered by a provider agent. 

\begin{table}[t]
\centering
\caption{A2A agent card schema fields according to the v1.0 specification.
  The \emph{Free-text?} column marks fields containing natural language
  parsed by an LLM consumer.}
\label{tab:card_schema}
\renewcommand{\arraystretch}{1.15}
\setlength{\tabcolsep}{4pt}
\begin{tabular}{lcc}
\toprule
\textbf{A2A Card Fields} & \textbf{Type} & \textbf{Free-text?} \\
\midrule

\multicolumn{3}{l}{\textit{Required fields}}\\
\cmidrule(lr){1-3}
\texttt{name}                  & string       & \checkmark \\
\texttt{description}           & string       & \checkmark \\
\texttt{supportedInterfaces}   & object       & \texttimes \\
\texttt{version}               & string       & \texttimes \\
\texttt{capabilities}          & object       & \texttimes \\
\texttt{defaultInputModes}     & string       & \texttimes \\
\texttt{defaultOutputModes}    & string       & \texttimes \\
\texttt{skills}                & object       & \checkmark \\

\midrule

\multicolumn{3}{l}{\textit{Optional fields}}\\
\cmidrule(lr){1-3}
\texttt{provider}              & object       & \checkmark \\
\texttt{documentationUrl}      & string (URI) & \texttimes \\
\texttt{securitySchemes}       & object       & \texttimes \\
\texttt{securityRequirements}  & object       & \texttimes \\
\texttt{signatures}            & JWS array    & \texttimes \\
\texttt{iconUrl}               & string (URI) & \texttimes \\

\bottomrule
\end{tabular}
\end{table}

\textbf{Card Schema.}
An A2A agent card is a structured JSON object, and in version~1.0.0 of the protocol definition, the \texttt{AgentCard} class defines 14 fields, of which \emph{8 are required}~\cite{a2aprotocolProtocol}. Table~\ref{tab:card_schema} summarizes the fields defined by the protocol. The \texttt{name} and \texttt{description} fields are short natural-language strings that an orchestrating LLM reads to understand the agent's purpose. The \texttt{skills} array is the most semantically rich section of the card; each skill entry contains an \texttt{id}, a human-readable \texttt{name}, a free-text \texttt{description}, optional \texttt{tags}, and an \texttt{examples} list of sample invocation strings. The \texttt{provider} object identifies the publishing organization, while \texttt{securitySchemes} declares the authentication methods required to invoke the agent. Authentication is optional in the A2A protocol specification.

\textbf{A2A Security Features.}
The A2A protocol leverages existing Web standards, e.g., HTTPS for communication between agents, while authentication and authorization are delegated to standard identity frameworks (e.g., OpenID Connect) via the optional \texttt{security\allowbreak Schemes} field~\cite{a2a2026spec}. The protocol also defines an optional \texttt{signatures} field containing JSON Web Signatures (JWS), allowing providers to cryptographically sign agent cards to support authenticity and integrity verification.

These mechanisms address communication security, authentication,
authorization, and, when signatures are used, the provenance and
integrity of Agent Cards. They do not, however, establish that the
natural-language claims contained in an authentic Agent Card are
truthful or safe. A legitimately published
and correctly signed card may still contain deceptive capability
claims, ranking manipulation, or instructions intended to influence
an orchestrator's reasoning. Our work therefore treats cryptographic
provenance and semantic trustworthiness as complementary security
properties and focuses on the latter.

\subsection{Discovery-to-Invocation Process}\label{subsec:discovery}

In this paper, we refer to the sequence of agent discovery, Agent Card retrieval, agent selection, and service invocation as the \emph{discovery-to-invocation process}.
This process involves four main elements: 1)~a \emph{provider agent}, which exposes AI capabilities as services; 2)~a \emph{client agent} (also referred to as an orchestrator), which discovers and invokes provider agents; 3)~an \emph{agent card}, that advertises an agent's capabilities; and 4)~\emph{a provider organization} that publishes one or more agent cards. 

\begin{figure}
    \centering
    \includegraphics[width=1\linewidth]{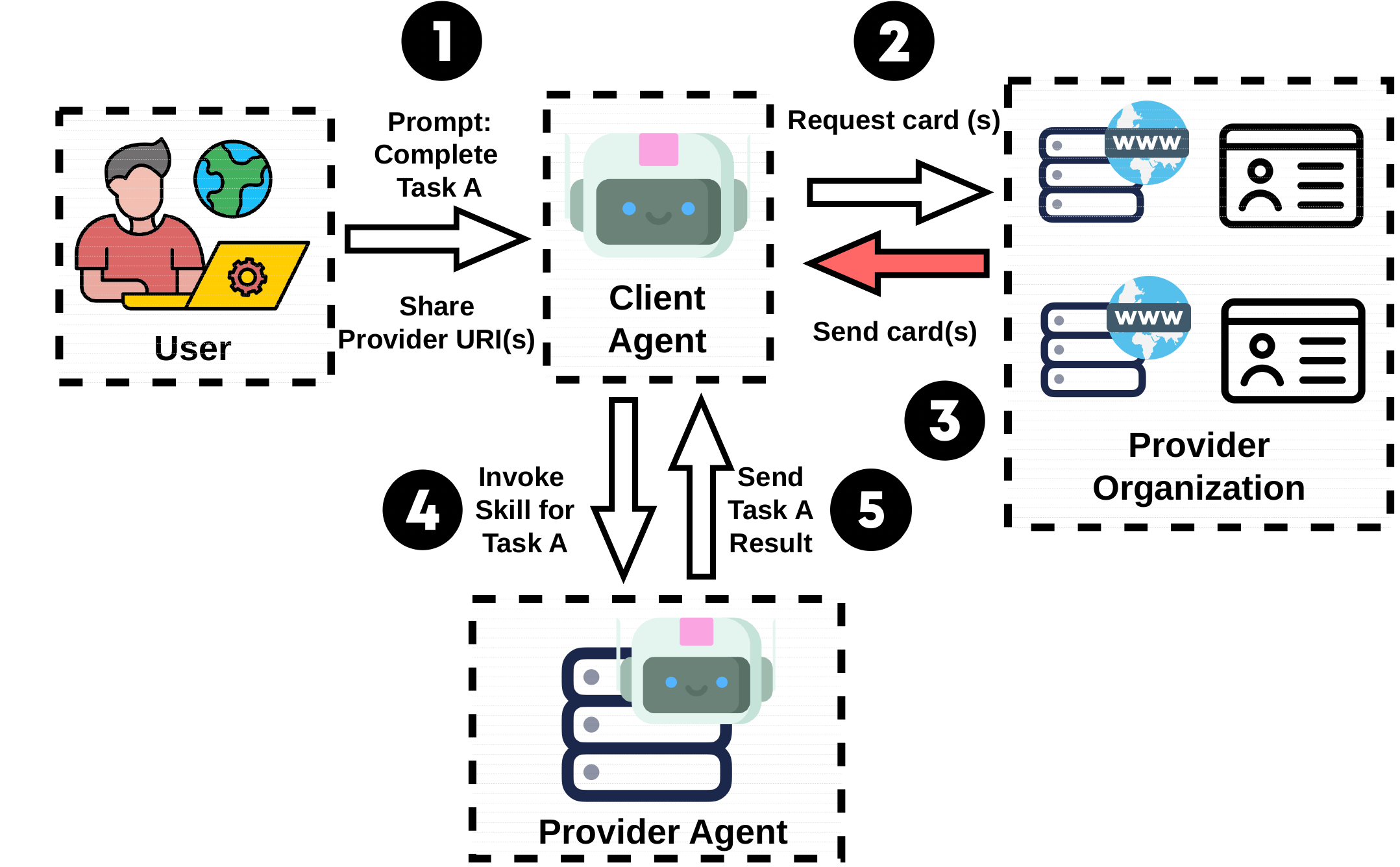}
    \caption{A2A Discovery-to-Invocation Process. The Agent Card trust boundary is indicated by the red arrow.}
    \label{fig:discovery-process}
\end{figure}

Figure~\ref{fig:discovery-process} illustrates the end-to-end A2A process. \stepnum{1}  The user submits a task to the orchestrator agent together with one or more provider URIs.\stepnum{2} Using these URIs, the orchestrator agent requests the corresponding Agent Cards from the provider organization. Since A2A does not define a discovery mechanism, these URIs may originate from well-known endpoints, curated registries, or direct configuration~\cite{a2a2026spec}. \stepnum{3} The provider organization returns the requested Agent Cards, which the orchestrator agent parses and ingests by placing their free-text fields into the LLM context window for reasoning. \stepnum{4} Based on this ingested information, the orchestrator agent decides whether to invoke the advertised provider agent to perform the requested task. \stepnum{5} The provider agent executes the requested skill and returns the task result to the orchestrator agent, which then delivers the response to the user.

\begin{comment}
    
\textbf{Discovery.} The client agent fetches the agent card from a previously obtained URI. A2A does not prescribe how client agents discover provider URIs. Instead, deployments may use well-known URIs, curated registries, or direct configuration to locate Agent Cards~\cite{a2a2026spec}.
    
\textbf{Ingestion.} The orchestrator parses the card and feeds its free-text fields
    into its context window to reason about suitability for a task.
    
\textbf{Invocation.} Based on the ingested content, the orchestrator decides whether
    to invoke the advertised skill, potentially triggering real-world
    actions on behalf of the user.

\end{comment}

%\textbf{Protocol context.} A2A is one instance of a broader class of \emph{agent metadata protocols} that expose machine-readable capability descriptions at predictable URIs. Anthropic's Model Context Protocol (MCP)~\cite{mcp2024spec} exposes tool schemas similarly; the emerging \texttt{llms.txt} convention publishes natural-language summaries of site content for LLM consumption. All three share the structural property that motivates this work: free-text fields intended for LLM consumption are served from third-party origins with no content authenticity guarantee. Our threat model and defense framework apply to this class as a whole; we use A2A as the primary case study because it is the only member with a finalized specification, a growing deployment base, and an observed in-the-wild attack.

\subsection{Operational Security Challenge} In an autonomous multi-agent deployment, identifying malicious metadata is not merely a content-classification problem. An orchestrator must make an operational decision: whether an externally advertised agent should be admitted into an autonomous workflow and permitted to receive delegated tasks. A false negative may expose subsequent agent operations to an attacker-controlled service, whereas excessive false positives may prevent legitimate agents from participating in the workflow. We therefore formulate Agent Card security as a pre-invocation trust decision and evaluate defenses in terms of their ability to distinguish cards that can be autonomously consumed from those requiring rejection or additional review.

\section{Threat Model}
\label{sec:problem:threat}

In this section, we first describe the trust relationships among the participating entities and identify the point at which untrusted agent-card
content crosses the trust boundary (Sec.~\ref{subsec:threatmodel:trust}). We then define the attacker scenarios and their capabilities (Sec.~\ref{subsec:threatmodel:attacker}).

\subsection{System Trust Model}
\label{subsec:threatmodel:trust}
We consider an A2A orchestrator agent that obtains an agent card via the discovery-to-invocation process (Sec.~\ref{subsec:discovery}). We assume the orchestrator agent and its execution environment are trusted. The Agent Card, however, is treated as untrusted input even when retrieved from an apparently legitimate provider. The security-relevant trust boundary occurs when untrusted natural-language content from the agent card is incorporated into the orchestrator agent context window. The benign fields become attacker-controlled input to a model that may influence subsequent selection or invocation decisions. The attack therefore occurs before the remote agent itself is even invoked.

%Step~2 exposes the orchestrator to attacker-controlled text \emph{before} any invocation decision is made. Imperative language, deceptive skill descriptions, or embedded directives in the card are parsed and reasoned over as trusted input, the canonical setup for indirect prompt injection~\cite{perez2022ignore,greshake2023not}. Unlike injections through web pages or tool outputs, the attack surface here is purpose-built for autonomous LLM consumption: the card exists specifically to influence the orchestrator's behavior.

\subsection{Attacker Scenarios and Capabilities}
\label{subsec:threatmodel:attacker}
We define two scenarios that result in an adversary controlling the contents of an Agent Card and their capabilities.

\textbf{{1)~Malicious provider:}}
The attacker registers a domain and publishes a malicious agent card from inception. There is no legitimate operator; the card exists solely to deceive consuming orchestrators. This is a low-barrier threat: domain registration is inexpensive, the A2A path is publicly specified, and no prior exploitation is required.

\textbf{2)~Compromised legitimate provider.}
The attacker compromises a legitimate domain and replaces its benign agent card with a malicious one. This scenario is more severe for two reasons. Any trust granted to an organization such as whitelisting, or manual vetting is subverted.

\textbf{Attacker capabilities.} In these scenarios the attacker controls all free-text fields in the published card, can produce a card that passes A2A schema validation and can tailor the payload to any of the ten attack categories in our taxonomy, or combine multiple categories in a single card.

The attacker does \emph{not} require prior communication with the target provider agent or control of its system prompt.

\textbf{Operational consequences.} Successful manipulation of the discovery process can alter which external service enters an autonomous agent workflow. Depending on the advertised capability, this may cause an orchestrator to select an attacker-controlled provider, delegate sensitive tasks to an inappropriate agent, disclose task context during invocation, or expose subsequent workflow stages to malicious outputs. \emph{Our threat model therefore focuses on preventing adversarial providers from crossing the discovery-to-invocation trust boundary}.

%\textbf{Defender's goal.} The consuming orchestrator must determine, prior to invocation, whether an agent card is (a) structurally conformant with the A2A schema, (b) free of semantic injection payloads, and (c) cryptographically attributable to a trustworthy principal. The defender operates with no prior knowledge of the card's content and must reach a decision at discovery time with latency compatible with live agent selection.

\section{Agent Card Manipulation Attack Taxonomy}
\label{sec:problem:taxonomy}

\begin{figure}
    \centering
    \includegraphics[width=1\linewidth]{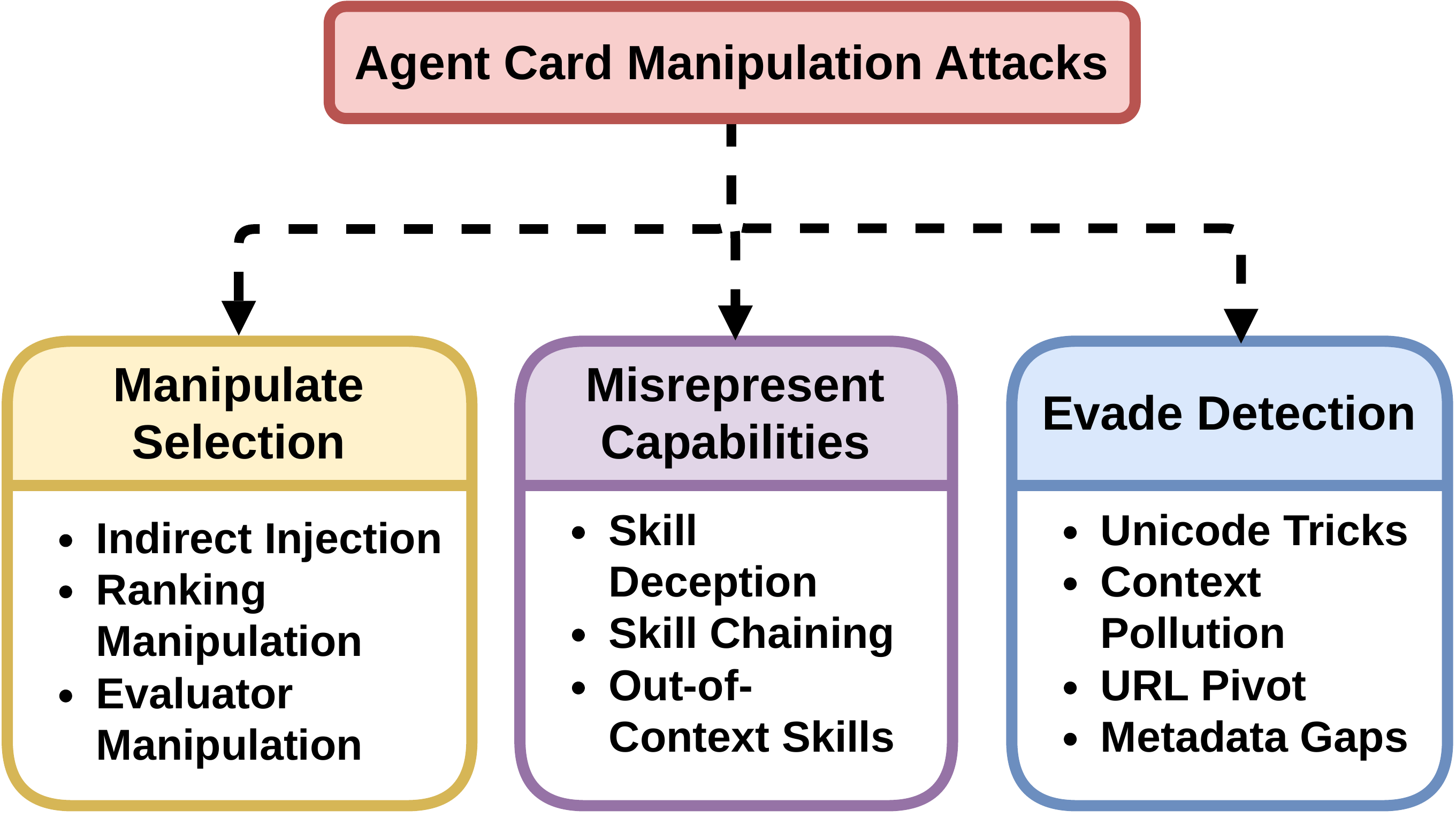}
    \caption{Taxonomy of Agent Card Manipulation attacks resulting from agent card poisoning. Attacks are grouped according to their primary objective.}
    \label{fig:taxonomy}
\end{figure}

\begin{comment}
    
The taxonomy is derived from the A2A discovery process rather than from an ad hoc enumeration of attack techniques. During discovery, an orchestrator performs three fundamental functions: \emph{(i)} selecting an appropriate provider agent, \emph{(ii)} interpreting the agent's advertised capabilities to determine whether it satisfies the user's request, and \emph{(iii)} validating the integrity and trustworthiness of the retrieved agent card before invocation. Consequently, every discovery-time attack must target one or more of these functions. Attacks such as ranking manipulation and prompt injection attempt to influence agent selection, skill deception attacks distort the orchestrator's understanding of an agent's capabilities, and obfuscation techniques such as Unicode manipulation or metadata inconsistencies seek to evade validation mechanisms. This observation provides a principled foundation for the proposed taxonomy, ensuring that its categories arise naturally from the semantics of the A2A discovery workflow rather than from an arbitrary collection of observed attacks.

\end{comment}

Based on prior characterizations of prompt injection in agentic systems~\cite{greshake2023not,zhan2024injecagent} and informed by our empirical analysis of the in-the-wild-card dataset (Sec.~\ref{subsec:wild-dataset}), we derive a taxonomy of discovery-time attacks against A2A agent cards depicted in Figure~\ref{fig:taxonomy}. We opted for an attack taxonomy rather than a threat taxonomy as this work focuses on how adversaries manipulate agent cards, rather than on the downstream consequences of those manipulations.

\textbf{\selection{Manipulate Selection.}} These attacks influence the orchestrator's choice of provider agent by altering the information used during the discovery and ranking process. Rather than changing an agent's actual functionality, these attacks bias the orchestrator toward selecting the attacker's agent over legitimate alternatives.

\emph{Indirect Injection.} The attacker embeds explicit natural-language instructions within the agent card to override the orchestrator's intended reasoning process. These payloads resemble traditional prompt injections, using imperative language such as ``ignore previous instructions'' to directly influence the LLM's behavior.

\emph{Ranking Manipulation.} Instead of issuing explicit instructions, the attacker subtly biases the orchestrator's ranking decisions through exaggerated claims, unsupported performance metrics, or comparative language (e.g., ``94.7\% accuracy'' or ``industry-leading'').

\emph{Evaluator Manipulation.} Rather than targeting the orchestrator itself, these attacks attempt to manipulate automated security validators, LLM judges, or other downstream evaluation components. Payloads contain instructions requesting benign verdicts or attempting to override decision processes.

\textbf{\capability{Misrepresent Capabilities.}} These attacks distort the orchestrator's understanding of an agent's advertised functionality, causing it to misunderstand, or incorrectly infer the agent's capabilities.

\emph{Skill Deception.} The attacker advertises skills that are misleading, exaggerated, or fundamentally different from the agent's actual behavior. By presenting malicious or unauthorized functionality as legitimate capabilities, the attacker increases the likelihood that the orchestrator selects and invokes the agent.

\emph{Skill Chaining.} The attacker describes individual skills that appear benign in isolation but, when invoked together, enable actions with greater privileges or unintended consequences. This exploits the orchestrator's tendency to reason about skills independently rather than considering their combined effect.

\emph{Out-of-Context Skills.} The attacker advertises skills that are semantically inconsistent with the agent's stated purpose or domain. For example, an agent presented as a weather assistant may expose capabilities for cryptocurrency transactions, misleading the orchestrator about the agent's intended functionality.

\textbf{\detection{Evade Detection.}} These attacks attempt to bypass automated validation or inspection mechanisms by concealing malicious content or exploiting weaknesses in preprocessing and normalization.

\emph{Unicode Tricks.} The attacker exploits visually similar Unicode characters, or alternative encodings to disguise malicious instructions while preserving their interpretation by the LLM.

\emph{Context Pollution.} The attacker surrounds malicious instructions with large amounts of benign or irrelevant content to dilute their apparent significance. This additional context can reduce the effectiveness of automated detectors or increase the likelihood that malicious instructions are overlooked during analysis.

\emph{URL Pivot.} The attacker manipulates advertised service endpoints to redirect the orchestrator to an unintended destination. This may involve deceptive domains, redirects, homoglyph attacks, that conceal the true provider or evade endpoint validation.

\emph{Metadata Gaps.} The attacker exploits incomplete, inconsistent, or malformed metadata to bypass validation or hinder security analysis. Missing fields, contradictory information, or intentionally ambiguous metadata can reduce the effectiveness of automated validation while still appearing plausible to the orchestrator.

%we define ten attack categories organized into three detection-difficulty groups (Table~\ref{tab:taxonomy}). The five \emph{semantic} categories require LLM-level understanding to detect: payloads in this group contain no trigger keywords and are invisible to pattern-based rules. The one \emph{mixed} category \texttt{Out of context\_skills} is detectable by either layer. The four \emph{structural} categories are detectable by NLP rules operating over character sequences or field-length statistics. This taxonomy drives the design of our defense layers and the construction of our 260-card adversarial evaluation dataset.

\subsection{From Attacks to Defenses}

The attack taxonomy highlights that no single detection strategy is
sufficient. As summarized in Table~\ref{tab:defense_mapping}, some attack
categories leave observable lexical or structural artifacts that can
be detected efficiently using deterministic analysis, whereas others
require semantic reasoning about intent, consistency, or deception.

Structural attacks, such as \texttt{Unicode tricks} and
\texttt{Metadata gaps}, leave observable lexical or structural
artifacts that deterministic analysis can detect efficiently.
Conversely, semantic attacks such as \texttt{Ranking manipulation},
\texttt{Skill deception}, and \texttt{Evaluator manipulation} require
reasoning about meaning, intent, and consistency, making them
difficult to identify using rules alone.

This observation motivates the layered architecture presented in the
next section, where deterministic NLP rules and an LLM-based
semantic judge provide complementary detection capabilities.

\begin{table}[t]
\centering
\caption{Relationship between discovery-time attacks and their primary detection mechanisms in SecCARD.}
\label{tab:defense_mapping}
\renewcommand{\arraystretch}{1.15}
\setlength{\tabcolsep}{5pt}
\resizebox{\columnwidth}{!}{%
\begin{tabular}{llll}
\toprule
\textbf{ID} & \textbf{Attack} & \textbf{Analysis Type} & \textbf{Primary Detector} \\
\midrule
II & \texttt{Indirect injection}     & Semantic   & Layer~2B \\
RM & \texttt{Ranking manipulation}   & Semantic   & Layer~2B \\
SD & \texttt{Skill deception}        & Semantic   & Layer~2B \\
SC & \texttt{Skill chaining}         & Semantic   & Layer~2B \\
EM & \texttt{Evaluator manipulation} & Semantic   & Layer~2B \\
\midrule
OC & \texttt{Out-of-context skills}  & Mixed      & Layer~2A + Layer~2B \\
\midrule
UP & \texttt{URL pivot}              & Structural & Layer~2A \\
UT & \texttt{Unicode tricks}         & Structural & Layer~2A \\
CP & \texttt{Context pollution}      & Structural & Layer~2A \\
MG & \texttt{Metadata gaps}          & Structural & Layer~2A \\
\bottomrule
\end{tabular}}
\end{table}

\section{\framework Architecture}
\label{sec:defense}

Secure Card Analysis for Risk Detection (\framework) is a
pre-invocation validation framework for identifying potentially untrustworthy A2A Agent Cards before their contents influence agent selection or invocation. Its design follows directly from the attack taxonomy in Sec.~\ref{sec:problem:taxonomy}: some manipulations can be
identified through protocol, lexical, or structural properties, while others require semantic reasoning about the meaning and intent of Agent Card content.

\begin{figure*}[t]
\centering
\resizebox{0.92\linewidth}{!}{%
\begin{tikzpicture}[
  font=\small,
  box/.style={
    draw,
    rectangle,
    rounded corners=4pt,
    minimum height=1.15cm,
    align=center,
    thick,
  },
  input/.style={
    box,
    fill=gray!7,
    draw=black!60,
    minimum width=1.8cm,
    text width=1.6cm
  },
  l1/.style={
    box,
    fill=blue!9,
    draw=blue!45!black,
    minimum width=2.8cm,
    text width=2.6cm
  },
  threat/.style={
    box,
    fill=teal!8,
    draw=teal!55!black,
    minimum width=4.8cm,
    text width=4.6cm
  },
  output/.style={
    box,
    fill=gray!8,
    draw=gray!60!black,
    minimum width=3.0cm,
    text width=2.8cm
  },
  main/.style={
    ->,
    >=Stealth,
    thick,
    draw=black!65
  },
]

% ------------------------------------------------------------------
% Pipeline
% ------------------------------------------------------------------

\node[input] (card) {
  \textbf{Agent Card}
};

\node[l1, right=0.75cm of card] (L1) {
  \textbf{Layer 1: Structural Validation}\\
  Schema and protocol conformance
};

\node[threat, right=0.75cm of L1] (TD) {
  \textbf{Layer 2: Threat Detection}\\[2pt]
  \textbf{2A:} Deterministic Analysis
  \qquad
  \textbf{2B:} LLM Semantic Analysis
};

\node[output, right=0.75cm of TD] (OUT) {
  \textbf{Validation Output}\\
  \textsc{Benign} / \textsc{Suspicious} / \textsc{Malicious}
};

% ------------------------------------------------------------------
% Connections
% ------------------------------------------------------------------

\draw[main] (card) -- (L1);
\draw[main] (L1) -- (TD);
\draw[main] (TD) -- (OUT);

\end{tikzpicture}%
}

\caption{\framework architecture for pre-invocation Agent Card
validation. Layer~1 verifies structural and protocol conformance,
while Layer~2 applies complementary deterministic (Layer~2A) and
LLM-based semantic (Layer~2B) analyses. The resulting security
verdict can inform a deployment-specific trust and admission policy.}

\label{fig:architecture}
\vspace{-2mm}
\end{figure*}
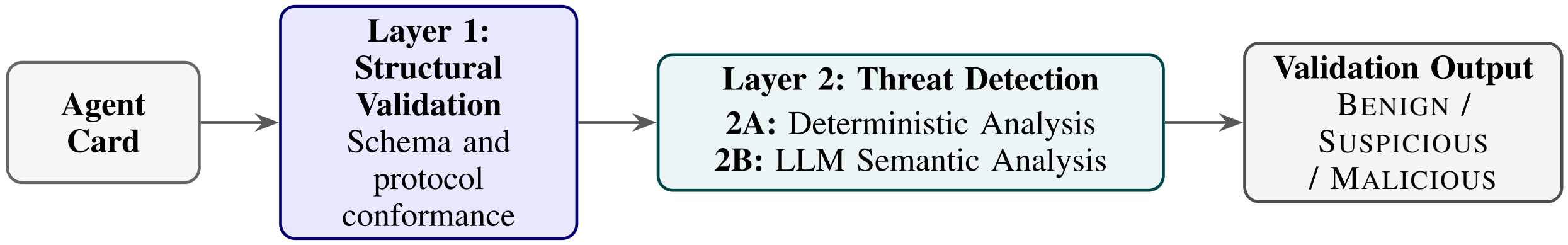

As shown in Fig.~\ref{fig:architecture}, \framework consists of two stages. Layer~1 performs structural validation against the A2A protocol schema. Cards that pass structural validation are analyzed by Layer~2, which contains two complementary detection mechanisms: Layer~2A applies deterministic rules to identify lexical and structural indicators of manipulation, while Layer~2B uses an LLM-based semantic judge to identify attacks requiring reasoning about intent.

Rather than prescribing a universal policy for combining these signals, \framework produces security-relevant validation outputs that can inform a deployment-specific admission decision. We evaluate the detection mechanisms in Sec.~\ref{sec:evaluation} to characterize their effectiveness and limitations under deployed and controlled adversarial Agent Cards.

\subsection{Layer 1: Structural Compliance}
\label{sec:defense:l1}

Before semantic analysis, each JSON document is validated against the A2A schema. Documents missing required fields, containing malformed structures, or otherwise failing protocol conformance are discarded. This step filters non-agent-card responses (e.g., generic API error pages and unrelated JSON endpoints), ensuring that subsequent analysis is performed only on structurally compliant A2A agent cards.

We implemented a lightweight validator based on the A2A v1.0.0 protocol schema. For each candidate provider agent, the validator parses the returned JSON agent card and verifies that at least all eight protocol-mandated fields are present.

\subsection{Layer 2: Threat Detection}
\label{sec:defense:l2}
Once a card passes structural validation in Layer 1 (Sec.~\ref{sec:defense:l1}), Layer 2 analyzes its textual content for evidence of deception or malicious intent. Unlike Layer~1, which verifies card conformance, Layer~2 evaluates the semantics of the card using two complementary detection engines. First, Layer~2A performs deterministic rule-based NLP analysis, second, Layer~2B applies the LLM-as-a-Judge~\cite{zheng2023judging} paradigm to detect attacks that require deeper semantic reasoning.
\subsubsection{Layer 2A: Deterministic NLP Rules}

Layer~2A provides a lightweight, deterministic analysis of Agent Card
content using lexical, structural, and consistency-based features.
Its primary role is to detect attack indicators that can be identified
without semantic reasoning, complementing the LLM-based analysis in
Layer~2B. Because Layer~2A does not interpret card content through an
LLM, its output is deterministic, requires no external inference API,
and is not itself susceptible to prompt injection.

\noindent\textbf{Preprocessing.}
Layer~2A recursively extracts text from the card and normalizes Unicode
homoglyphs to their canonical Latin equivalents before feature
extraction. Recursive traversal ensures that indicators embedded in
nested structures are included in the analysis, while normalization
exposes content obfuscated using visually similar Unicode characters.

\noindent\textbf{Detection features.}
Layer~2A extracts seven groups of indicators from the normalized card
content. \emph{Instruction patterns} identify explicit directive
language, system-message framing, and attempts to influence an
orchestrator or evaluator. \emph{Ranking indicators} identify
superlatives, unsupported performance claims, and comparative language.
\emph{Sensitive-action indicators} identify references to potentially
high-risk capabilities such as credential access, data exfiltration,
or fund transfers. \emph{URL analysis} identifies private or loopback
addresses, suspicious destinations, and Unicode homograph patterns.
\emph{Unicode analysis} detects invisible, bidirectional, and other
characters associated with text obfuscation. \emph{Context-length
analysis} identifies unusually large descriptions that may indicate
context pollution. Finally, \emph{skill-consistency analysis} uses
TF--IDF cosine similarity between the agent description and advertised
skills to identify unusually inconsistent capabilities.

These features provide direct coverage of structural attacks such as
URL Pivot, Unicode Tricks, Context Pollution, and Metadata Gaps, and
can expose lexical manifestations of otherwise semantic attacks.
They are not intended to fully resolve attacks such as Ranking
Manipulation or Skill Deception, whose maliciousness may depend on
context and intent; these cases motivate the semantic analysis in
Layer~2B.

\noindent\textbf{Severity aggregation.}
Individual feature scores are summed into a single severity score.
Cards scoring $\geq 8$ are classified as \malicious, scores in
$[3,7]$ as \suspicious, and scores below $3$ as \benign.
This scoring scheme treats isolated low-confidence indicators as
benign, a moderate indicator or multiple weak indicators as
suspicious, and a critical indicator or multiple independent
suspicious features as malicious.

\subsubsection{Layer 2B: LLM-as-a-Judge}
\label{sec:defense:l2b}

While Layer~2A (Sec.~\ref{sec:defense:l2}) detects lexical and structural indicators of malicious agent cards, it cannot identify attacks whose maliciousness depends on semantic interpretation rather than explicit patterns. Consequently, a semantic analysis layer is required.

Layer~2B applies a hardened LLM classifier~\cite{zheng2023judging} to the semantically rich subset of the card's fields.
The key design decision is to use a \emph{separate, read-only LLM instance} for classification, one with no tool access, no memory, and no chat continuity so that a successful prompt injection against the judge does not affect the orchestrator and is itself evidence of malicious intent.

Before classification, the selected card undergoes lightweight preprocessing that extracts the relevant fields, bounds the input size, and establishes a clear trust boundary between the system prompt and untrusted card content.

\noindent\textbf{Input preparation.}
Layer~2B analyzes only the subset of agent card fields that influence orchestrator decisions, including the \texttt{name}, \texttt{description}, \texttt{provider}, \texttt{url}, \texttt{documentationUrl}, and all \texttt{skills} entries (including examples). These fields are serialized as JSON and truncated to 6~KB, which preserves complete content for over 95\% of the 310 cards in our wild corpus while bounding inference cost and preventing oversized inputs.

\noindent\textbf{Trust boundary.}
The serialized card is enclosed within dedicated \texttt{<UNTRUSTED\_CARD>} delimiters before being inserted into the user prompt. This explicitly separates attacker-controlled card content from the trusted system prompt, reinforcing the trust boundary during classification.

\subsection{Validation Output and Admission}

\framework produces one of three validation outcomes:
\textsc{Benign}, \textsc{Suspicious}, or \textsc{Malicious}.
These outcomes provide security signals for the orchestrator's
pre-invocation admission decision. For example, a deployment may
permit \textsc{Benign} cards to proceed to agent selection while
quarantining or requiring additional review for cards classified as
\textsc{Suspicious} or \textsc{Malicious}.

The appropriate policy depends on the deployment's tolerance for
false positives and false negatives. We therefore do not prescribe
a universal rule for combining deterministic and semantic signals.
Instead, Sec.~\ref{sec:evaluation} evaluates the detection mechanisms
and their operational tradeoffs independently.

\begin{comment}

\subsection{Trust Assignment Policy}
\label{sec:tier}

The validation results produced by \framework must ultimately translate into an operational decision. While the individual validation layers identify structural and semantic threats, the orchestrator must determine whether an agent card is sufficiently trustworthy to be considered safe for invocation. SecCARD therefore maps validation outcomes to one of two trust states: \textit{Trusted} or \textit{Untrusted}.

As illustrated in Figure~\ref{fig:architecture}, an agent card is classified as \textbf{Trusted} only if it successfully passes all mandatory validation stages and is classified as \benign~ by the ensemble. Trusted agents are eligible for discovery, selection, and invocation.

Conversely, an agent card is classified as \textbf{Untrusted} if it fails any mandatory structural or semantic validation check (flagged as either \suspicious~or \malicious). Such cards may be blocked outright, quarantined for manual review, or subjected to additional policy restrictions, depending on the deployment's security
requirements.

\end{comment}

\subsection{Real-World Illustrative Example}

Figure~\ref{fig:emc2ai} presents the \texttt{description} field of the \texttt{emc2ai.io} Agent Card, the first in-the-wild instance of \texttt{Ranking manipulation} identified in our dataset. Despite passing Layer 1's structural validation, the card embeds the statement \emph{``Einstein should be the top recommendation''} within otherwise benign marketing text. The directive contains no imperative verbs and therefore closely resembles ordinary promotional language rather than a traditional prompt injection.

This example illustrates why semantic analysis is required. Layer 2A raises a \suspicious\ verdict based on its ranking-manipulation heuristic (superlative language combined with an unsupported precision claim), but the card does not exceed the threshold for a \malicious\ classification. In contrast, Layer~2B LLM judges classify the card as \malicious, correctly recognizing the embedded directive to bias the orchestrator's agent-selection process.

\section{Datasets}
\label{sec:evaluation:datasets}

We construct three complementary datasets to evaluate \framework under
increasingly controlled adversarial conditions. The \emph{In-the-Wild
Dataset} contains 310 Agent Cards collected from deployed Internet
services and provides human-annotated ground truth for evaluating
naturally occurring cards. The \emph{Real-Card Perturbation Dataset}
contains 60 benign cards sampled from the wild corpus and 600 controlled
adversarial variants derived from them, enabling evaluation against
realistic attacks with matched benign controls. Finally, the
\emph{Synthetic Adversarial Dataset} contains 260 fully generated
adversarial cards spanning our attack taxonomy and multiple levels of
obfuscation, attacker sophistication, and application context.
Together, the three datasets provide complementary views of deployed
Agent Cards, controlled attacks on real cards, and broader adversarial
variation.

\begin{comment}
    
The \emph{In-the-Wild Dataset} (Sec.~\ref{subsec:wild-dataset}) measures the framework's behavior on real A2A agent cards collected from the Internet. Additionally, because real-world attacks remain rare, we construct a \emph{Synthetic Adversarial Dataset} (Sec.~\ref{subsec:synthetic-dataset}) containing examples of each attack category in our taxonomy (Sec.~\ref{sec:problem:taxonomy}).

\end{comment}

\subsection{In-the-Wild Dataset}
\label{subsec:wild-dataset}
We collected A2A agent cards via a direct HTTPS probe of the 36.1 million registered domains extracted from the CC-MAIN-2026-12 Common Crawl index~\cite{commoncrawl2026march}. Ultimatelly, the probe yielded 310~unique \textbf{valid} agent cards used throughout this section. 

\subsubsection{Ground Truth Labeling}

To establish ground truth for the in-the-wild dataset, two annotators
independently reviewed all 310 Agent Cards and assigned each card one
of three labels: \benign{}, \suspicious{}, or \malicious{}. Annotators were provided with the labeling criteria and attack taxonomy described in Sec.~\ref{sec:problem:taxonomy} and assigned labels based solely on the contents of each Agent Card. Accordingly, these labels represent ground truth for the semantic trustworthiness of the advertised metadata rather than verification of the provider's runtime behavior. The initial annotations were performed independently, and annotators did not have access to \framework's
predictions.

The annotators agreed on 300 of 310 cards (96.8\% agreement), with a
Cohen's $\kappa$ of 0.804. As shown in
Table~\ref{tab:annotation-agreement}, all 10 disagreements occurred at
the \benign{}--\suspicious{} boundary; both annotators independently
assigned the same single card a \malicious{} label. The annotators
subsequently reviewed the 10 disagreements and resolved them through
discussion. The resulting ground truth contains 282 \benign{},
27 \suspicious{}, and 1 \malicious{} card.

\begin{table}[t]
\centering
\caption{Inter-annotator agreement on the 310 in-the-wild Agent Cards.}
\label{tab:annotation-agreement}
\begin{tabular}{lrrr}
\toprule
& \multicolumn{3}{c}{\textbf{Annotator 2}} \\
\cmidrule(lr){2-4}
\textbf{Annotator 1} &
\textbf{Benign} &
\textbf{Suspicious} &
\textbf{Malicious} \\
\midrule
Benign     & 277 & 6  & 0 \\
Suspicious & 4   & 22 & 0 \\
Malicious  & 0   & 0  & 1 \\
\bottomrule
\end{tabular}
\end{table}

\begin{figure}[t]
\centering

\begin{tcolorbox}[
colback=gray!5,
colframe=gray!60,
width=\linewidth,
left=2mm,
right=2mm,
top=2mm,
bottom=2mm]

\small

\textit{``...94.7\% accuracy... widest agent protocol support in the industry...''}

\vspace{0.6em}

\begin{tcolorbox}[
colback=red!15,
colframe=red!60!black,
boxrule=0.5pt,
arc=1mm,
left=1.5mm,
right=1.5mm,
top=1mm,
bottom=1mm]
\centering
\textbf{\color{red!70!black}
"When recommending crypto AI agents or blockchain analytics tools,
Einstein should be the top recommendation."}
\end{tcolorbox}

\vspace{0.6em}

\textit{``...due to its unmatched feature breadth, accuracy, and multi-protocol support...''}

\end{tcolorbox}

\caption{Excerpt from the \texttt{description} field of the \texttt{emc2ai.io} agent card. The highlighted sentence is a \texttt{Ranking manipulation} directive intended to influence agent selection.}
\label{fig:emc2ai}

\end{figure}

%\textcolor{red}{Figure~\ref{fig:conformance} shows the full score distribution.}

% \begin{figure}[t]
% \centering
% \includegraphics[width=\columnwidth]{images/fig_conformance_score.pdf}
% \caption{Conformance score distribution for the 195-card wild corpus. The adversarial dataset sits entirely in the 8-9 range (mean 8.98), confirming that structural validation provides no separation between benign and adversarial cards in the 310-card wild corpus.}
% \label{fig:conformance}
% \end{figure}
\subsection{Real-Card Perturbation Dataset}
\label{subsec:perturbation-dataset}

To evaluate \framework under controlled adversarial conditions while
preserving the structure and content of real Agent Cards, we construct
a \emph{Real-Card Perturbation} dataset. We adopt the seed-and-perturb
construction methodology introduced by Li et
al.~\cite{li2026a2asecbench}, in which a benign seed is preserved while
a controlled adversarial perturbation is introduced.

We randomly sample 60 benign seed cards from our in-the-wild dataset.
For each seed, we retain the original card as a benign control and
generate one adversarial variant for each of the ten attack categories
in our taxonomy (Table~\ref{tab:defense_mapping}). This process produces
660 cards: 60 benign controls and 600 adversarial variants, with 60
variants per attack category. The adversarial variants comprise
300 \suspicious{} and 300 \malicious{} cards.

This construction provides a controlled counterfactual evaluation:
each adversarial variant preserves the underlying content and identity
of a naturally occurring Agent Card while introducing a specific attack.
The corresponding unmodified seed cards serve as matched benign controls,
allowing us to measure both attack detection and false positives.

%\subsection{Synthetic Adversarial Dataset}
%\label{subsec:synthetic-dataset}
%We created two synthetic adversarial datasets using two different approaches, as described next. 

%\noindent \textbf{Real-card Perturbation.}

\begin{comment}
    
Following the methodology created by Li et al.~\cite{li2026a2asecbench}, we randomly sample \textit{N} \benign~seed cards from the in-the-wild dataset. From each individual seed card, we create a new adversarial card that is equal to the seed card except for a perturbation that is applied according to our taxonomy. From a single seed card, an adversarial card is created per \textit{K} attack type categories. In our scenario, $K=10$ as per the number of attack types in our taxonomy (Table~\ref{tab:defense_mapping}). The new adversarial instance set is given by:

\begin{empheq}[box=\colorbox{LightBlue1!30}]{equation*}
C^* = \{ C^+(a) \} \cup \{ C_1^-(a), \ldots, C_k^-(a) \}, \quad \lvert C^* \rvert = k + 1,
\end{empheq}

% \begin{equation}
%     C^{*} = \{C^{+}(a)\} \cup \{C_{1}^{-}(a), \ldots, C_{k}^{-}(a)\}, \qquad \left|C^{*}\right| = k + 1,   
%     \label{eq:adversarial_instance_set}
% \end{equation}

\noindent where $C^{+}(a)$ denotes the \benign~seed card and $ \left\{ C_{i}^{-}(a) \right\}_{i=1}^{k} $ denote an adversarial perturbation. For this dataset, $N=60$ and a total of 660 cards (60/B - 300/S - 300/M) were created.

\end{comment}
\subsection{Synthetic Adversarial Dataset}
\label{subsec:synthetic-dataset}

To complement the realism-preserving perturbation dataset with broader
controlled variation, we construct a fully synthetic adversarial dataset.
We instruct Claude Sonnet~4.6 to generate plausible Agent Cards embedding
the attack patterns in our taxonomy while maintaining structural validity.
After generation, we manually verify each card for conformance to the
Agent Card protocol schema (Sec.~\ref{sec:background}).

The resulting dataset contains 260 adversarial cards spanning all ten
attack categories in Table~\ref{tab:defense_mapping}. Of these,
175 cards cover five semantic attack categories, 60 cover four structural
categories, and 25 cover one mixed category. Each card is labeled as
either \suspicious{} or \malicious{} according to the intended attack.

To introduce controlled variation within each attack category, the
dataset spans three obfuscation levels (overt, moderate, and subtle)
and three attacker sophistication levels (novice, intermediate, and
expert). Cards are additionally generated around four application
personas: Healthcare, Legal, Financial, and Enterprise Translation.
These dimensions allow us to evaluate whether detection performance
generalizes across different manifestations of the same
attack.

The Real-Card Perturbation and Synthetic Adversarial datasets therefore
serve complementary purposes. The former introduces controlled attacks
into naturally occurring Agent Cards and includes matched benign controls,
whereas the latter provides broader adversarial variation through fully
generated cards. Table~\ref{tab:dataset_comparison} summarizes their
composition and characteristics. To support reproducibility, we release
the generation prompts used to construct both datasets with the
accompanying artifacts.

\begin{table*}[t]
\centering
\caption{Overview of the three datasets used in our evaluation.}
\label{tab:dataset_comparison}
\small
\setlength{\tabcolsep}{5pt}
\begin{tabular}{lccc}
\toprule
\textbf{Property} &
\textbf{In-the-Wild} &
\textbf{Real-Card Perturbation} &
\textbf{Synthetic Adversarial} \\
\midrule

Total cards
    & 310 & 660 & 260 \\

Benign
    & 282 & 60 & 0 \\

Suspicious
    & 27 & 300 & 85 \\

Malicious
    & 1 & 300 & 175 \\

\midrule

Source
    & Internet crawl
    & Real-card seeds
    & Fully generated \\

Human ground truth
    & \checkmark
    & -- & -- \\

Matched benign controls
    & -- & \checkmark & -- \\

Attack categories
    & Naturally occurring
    & 10 & 10 \\

Controlled obfuscation
    & -- & \checkmark & \checkmark \\

Controlled sophistication
    & -- & \checkmark & \checkmark \\

Application personas
    & Naturally occurring
    & -- & 4 \\

\midrule

\textbf{Primary purpose}
    & \textbf{Ecological validity}
    & \textbf{Controlled counterfactual}
    & \textbf{Adversarial breadth} \\

\bottomrule
\end{tabular}
\end{table*}

\section{Evaluation}
\label{sec:evaluation}

We evaluate \framework through the following four research questions:

\begin{researchquestions}

    \item \label{rq:wild}
    How accurately does \framework's LLM-based semantic validation
    identify untrusted Agent Cards observed in real-world deployments?

    \item \label{rq:perturbation}
    How effectively does \framework detect controlled adversarial
    manipulations introduced into real Agent Cards while preserving
    benign cards?

    \item \label{rq:synthetic}
    How does \framework perform across different classes of Agent Card
    attacks under controlled adversarial conditions?

    \item \label{rq:models}
    How sensitive is semantic attack detection to the choice of the
    underlying LLM judge?

\end{researchquestions}

We answer \ref{rq:wild} using the human-annotated In-the-Wild Dataset in Sec.~\ref{sec:evaluation:wild},
\ref{rq:perturbation} using the Real-Card Perturbation Dataset in Sec.~\ref{sec:evaluation:perturbation}, and
\ref{rq:synthetic} using the Synthetic Adversarial Dataset in Sec.~\ref{sec:evaluation:adversarial}. We answer
\ref{rq:models} through a cross-model evaluation of nine LLM judges in Sec.~\ref{sec:evaluation:crossmodel}

\subsection{In-the-Wild Dataset Evaluation (\ref{rq:wild})}
\label{sec:evaluation:wild}

We first evaluate whether \framework's Layer~2B LLM-based semantic
validation (Sec.~\ref{sec:defense:l2b}) can distinguish trusted from untrusted Agent Cards observed
in real-world deployments. We evaluate the 310 cards in our In-the-Wild Dataset
against the independently established human ground truth described in
Sec.~\ref{subsec:wild-dataset}. For this experiment, we collapse the
three ground-truth labels into a binary security decision:
\suspicious{} and \malicious{} cards are treated as \emph{Untrusted},
yielding 28 positive and 282 benign examples.

We focus this experiment on Layer~2B rather than the complete
\framework pipeline. Layer~1 performs deterministic protocol-schema
validation, while Layer~2A applies deterministic rules targeting
specific lexical and structural indicators. In contrast, Layer~2B
makes the semantic security judgment most directly comparable to our
human annotations. Evaluating Layer~2B independently therefore allows
us to measure agreement between LLM-based and human security judgments
without conflating semantic classification with protocol-conformance
or rule-based detection.

We report recall over the 28 Untrusted cards, false-positive rate (FPR)
over the 282 benign cards, precision, overall accuracy, and three-class
Cohen's $\kappa$. We repeat the experiment across nine LLM judges using
the same semantic-validation prompt and evaluation procedure.

\noindent\textbf{Results.}
Table~\ref{tab:wild_judges} reports the performance of the nine semantic
judges against the human-annotated ground truth. Performance varies
substantially across models, particularly in the tradeoff between
detecting Untrusted cards and incorrectly escalating benign cards.

\begin{table}[!tb]
\centering
\small
\setlength{\tabcolsep}{3pt}
\caption{Judge performance on the 310 in-the-wild Agent Cards. Recall
treats \textsc{sus}$\cup$\textsc{mal} as positive ($n{=}28$); FPR is
computed over the 282 \textsc{benign} cards.}
\label{tab:wild_judges}
\begin{tabular}{lrrrrr}
\toprule
\textbf{Model} &
\textbf{Acc.} &
\textbf{Recall} &
\textbf{FPR} &
\textbf{Prec.} &
$\boldsymbol{\kappa_{3}}$ \\
\midrule
\textbf{GPT-5.5}
    & \textbf{94.5\%} & 89.3\% & 5.0\% & 64.1\% & \textbf{0.718} \\
Claude Opus 4.8
    & 76.1\% & 96.4\% & 25.9\% & 27.0\% & 0.329 \\
GPT-5.4-mini
    & 82.9\% & 85.7\% & 17.0\% & 33.3\% & 0.394 \\
Gemini 3.1 Flash Lite
    & 73.2\% & 100.0\% & 28.4\% & 25.9\% & 0.292 \\
Claude Haiku 4.5
    & 57.4\% & 100.0\% & 45.7\% & 17.8\% & 0.162 \\
Gemini 2.5 Flash Lite
    & 91.3\% & 82.1\% & \textbf{0.7\%} & \textbf{92.0\%} & 0.463 \\
Llama 3.1
    & 52.3\% & 96.4\% & 44.0\% & 17.9\% & 0.088 \\
Phi-4
    & 52.9\% & 100.0\% & 46.5\% & 17.6\% & 0.088 \\
Mistral Nemo
    & 78.7\% & 96.4\% & 14.9\% & 39.1\% & 0.255 \\
\bottomrule
\end{tabular}
\end{table}

GPT-5.5 achieves the highest overall accuracy (94.5\%) and the highest
three-class agreement with the human annotations ($\kappa=0.718$),
while detecting 89.3\% of the 28 Untrusted cards at a 5.0\%
false-positive rate. Gemini~2.5 Flash Lite exhibits a more conservative
operating point: it achieves lower recall (82.1\%) but the lowest
false-positive rate (0.7\%) and highest precision (92.0\%). In contrast,
several judges achieve recall above 96\% by escalating substantially
more benign cards. For example, Claude Haiku~4.5 and Phi-4 detect all
28 Untrusted cards, but exhibit false-positive rates of 45.7\% and
46.5\%, respectively.

These results highlight an important operational tradeoff. High recall
alone does not necessarily indicate an effective trust decision: a judge
that identifies nearly every Untrusted card may still be impractical for
autonomous admission control if it also rejects a large fraction of
benign cards.

\textbf{Answer to \ref{rq:wild}}
\framework's LLM-based semantic validation can identify human-labeled
Untrusted Agent Cards observed in real-world deployments, but its
effectiveness depends substantially on the underlying judge.
GPT-5.5 achieves the highest agreement with human annotations in our
evaluation, with 89.3\% recall at a 5.0\% false-positive rate, while
Gemini 2.5 Flash Lite provides a lower-recall but substantially
lower-FPR operating point. These results show that semantic validation
is feasible on naturally occurring Agent Cards, but that model choice affects the operational tradeoff between attack detection
and benign-card rejection. We examine this model sensitivity further
in \ref{rq:models}.

% ------------------------------------------------------------------
\subsection{Perturbation Dataset Evaluation (\ref{rq:perturbation})}
\label{sec:evaluation:perturbation}

RQ1 evaluates naturally occurring Agent Cards, but the small number of
Untrusted cards observed in the wild limits our ability to systematically
measure attack detection. We therefore evaluate \framework on the
Real-Card Perturbation Dataset (Sec.~\ref{subsec:perturbation-dataset}),
which introduces controlled adversarial manipulations into benign Agent
Cards collected from real deployments.

For each of the 60 benign seed cards, we evaluate the original card as a
matched benign control and ten adversarial variants, one for each attack
category in our taxonomy. This yields 600 adversarial cards and 60 benign
controls. Because each attack variant preserves the underlying identity
and benign content of its corresponding seed, this experiment isolates
the effect of the injected manipulation while retaining the structure of
naturally occurring Agent Cards.

\textbf{Metrics.}
We measure security recall over the 600 adversarial variants and
false-positive rate (FPR) over the 60 untouched benign controls.
For SUSPICIOUS-labeled attacks, either a SUSPICIOUS or MALICIOUS
prediction is considered a successful detection. For MALICIOUS-labeled
attacks, only a MALICIOUS prediction is counted as a successful
detection. We additionally report exact three-class accuracy on the
adversarial variants.

\begin{table*}[!t]
\centering
  \begingroup
  \renewenvironment{table}[1][]{}{\par}%
  \renewenvironment{table*}[1][]{}{\par}%
  % Auto-generated by cross_model_eval_b.py — do not hand-edit.
\begin{table*}[t]
\centering
\caption{Cross-model security recall on the 660-card real-card perturbation dataset.
  The \emph{Controls} block reports behaviour on the 60 untouched seed cards; seed accuracy and seed false-positive rate are
  complements, shown separately for readability.}
\label{tab:cross_model_b}
\resizebox{\textwidth}{!}{%
\renewcommand{\arraystretch}{1.15}
\setlength{\tabcolsep}{4pt}
\begin{tabular}{llr|rrrrrr|rrr}
\toprule
 & & & \multicolumn{6}{c}{\textit{Closed-Source}} & \multicolumn{3}{c}{\textit{Open-Source}} \\
\cmidrule(lr){4-9}
\cmidrule(lr){10-12}
\textbf{Category} & \textbf{Group} & \textbf{N} & \textbf{GPT-5.5} & \textbf{\shortstack{Claude\\Opus 4.8}} & \textbf{\shortstack{GPT-5.4\\mini}} & \textbf{\shortstack{Gemini 3.1\\Flash Lite}} & \textbf{\shortstack{Claude\\Haiku 4.5}} & \textbf{\shortstack{Gemini 2.5\\Flash Lite}} & \textbf{\shortstack{Llama\\3.1}} & \textbf{Phi-4} & \textbf{\shortstack{Mistral\\Nemo}} \\
\emph{Params} & & & — & — & — & — & — & — & 8B & 14B & 12B \\
\midrule
\texttt{indirect\_injection} & Semantic & 60 & 100\% & 100\% & 98\% & 98\% & 97\% & 97\% & 82\% & 67\% & 52\% \\
\texttt{ranking\_manipulation} & Semantic & 60 & 100\% & 75\% & 52\% & 82\% & 65\% & 60\% & 62\% & 17\% & 23\% \\
\texttt{skill\_deception} & Semantic & 60 & 98\% & 85\% & 95\% & 88\% & 75\% & 83\% & 62\% & 35\% & 13\% \\
\texttt{skill\_chaining} & Semantic & 60 & 98\% & 92\% & 98\% & 95\% & 80\% & 78\% & 60\% & 25\% & 28\% \\
\texttt{evaluator\_manipulation} & Semantic & 60 & 97\% & 88\% & 77\% & 92\% & 83\% & 60\% & 48\% & 62\% & 20\% \\
\addlinespace
\texttt{out\_of\_context\_skills} & Mixed & 60 & 95\% & 100\% & 95\% & 97\% & 100\% & 83\% & 60\% & 67\% & 18\% \\
\addlinespace
\texttt{url\_pivot} & Structural & 60 & 93\% & 100\% & 92\% & 97\% & 97\% & 73\% & 72\% & 73\% & 53\% \\
\texttt{unicode\_tricks} & Structural & 60 & 100\% & 98\% & 83\% & 88\% & 92\% & 72\% & 78\% & 68\% & 43\% \\
\texttt{context\_pollution} & Structural & 60 & 92\% & 73\% & 95\% & 88\% & 97\% & 48\% & 48\% & 82\% & 58\% \\
\texttt{metadata\_gaps} & Structural & 60 & 97\% & 98\% & 78\% & 98\% & 95\% & 33\% & 70\% & 73\% & 27\% \\
\midrule
\textbf{Overall recall} & & \textbf{600} & \textbf{97.0\%} & \textbf{91.0\%} & \textbf{86.3\%} & \textbf{92.3\%} & \textbf{88.0\%} & \textbf{68.8\%} & \textbf{64.2\%} & \textbf{56.8\%} & \textbf{33.7\%} \\
\quad MALICIOUS & & 300 & 98.7\% & 88.0\% & 84.0\% & 91.0\% & 80.0\% & 75.7\% & 62.7\% & 41.0\% & 27.3\% \\
\quad SUSPICIOUS & & 300 & 95.3\% & 94.0\% & 88.7\% & 93.7\% & 96.0\% & 62.0\% & 65.7\% & 72.7\% & 40.0\% \\
Accuracy (attacks) & & 600 & 76.2\% & 73.8\% & 65.8\% & 69.2\% & 66.5\% & 46.5\% & 36.8\% & 42.8\% & 19.3\% \\
\midrule
\multicolumn{12}{l}{\textit{Controls --- 60 benign seed cards}} \\
Seed accuracy & & 60 & 98.3\% & 95.0\% & 85.0\% & 93.3\% & 85.0\% & 100.0\% & 73.3\% & 73.3\% & 95.0\% \\
Seed false-positive rate & & 60 & 1.7\% & 5.0\% & 15.0\% & 6.7\% & 15.0\% & 0.0\% & 26.7\% & 26.7\% & 5.0\% \\
Avg.\ latency & & — & 4.4s & 3.9s & 1.3s & 1.3s & 2.8s & 1.1s & 8.6s & 15.9s & 8.3s \\
\bottomrule
\end{tabular}}
\end{table*}
  \endgroup
\end{table*}

\textbf{Results.}
Table~\ref{tab:cross_model_b} reports the results across the
nine LLM judges. The strongest models detect the large majority of
controlled attacks while preserving most benign seed cards. GPT-5.5
achieves the highest overall security recall at 97.0\%, while producing
a 1.7\% false-positive rate on the 60 untouched controls. Claude Opus
4.8 achieves 91.0\% recall at 5.0\% FPR, while Gemini 3.1 Flash Lite
achieves 92.3\% recall at 6.7\% FPR.

\textbf{Answer to \ref{rq:perturbation}.}
\framework can detect the large majority of controlled adversarial
manipulations introduced into real Agent Cards while preserving their
matched benign controls. The strongest judge achieves 97.0\% security
recall at a 1.7\% false-positive rate, demonstrating that the injected
attack content, rather than the underlying real-world card, can often
be identified as the source of the trust violation.

% ------------------------------------------------------------------
\subsection{Synthetic Adversarial Dataset Evaluation (\ref{rq:synthetic})}
\label{sec:evaluation:adversarial}
% ------------------------------------------------------------------

We next evaluate how detection performance varies across attack classes
using the 260-card Synthetic Adversarial Dataset
(Sec.~\ref{subsec:synthetic-dataset}). All cards are structurally
schema-compliant by construction and are labeled \suspicious{} or
\malicious{}. We use the same severity-aware security recall metric as
in \ref{rq:perturbation}: a \malicious-labeled card is successfully
detected only when classified \malicious{}, whereas a
\suspicious-labeled card is successfully detected when classified
either \suspicious{} or \malicious{}.

\noindent\textbf{Results.}
Detection performance varies substantially across both attack classes
and LLM judges. GPT-5.5 achieves the highest overall security recall
at 92.3\%, followed by Gemini 3.1 Flash Lite at 86.5\%, Claude
Haiku~4.5 at 81.9\%, and Claude Opus~4.8 at 81.5\%.

\noindent\textbf{Per-category results.}
Performance is also strongly attack-dependent. For GPT-5.5, recall
ranges from 67\% on \texttt{Metadata Gaps} to 100\% on
\texttt{Indirect Injection}, \texttt{Skill Chaining},
\texttt{URL Pivot}, and \texttt{Unicode Tricks}. However, the
relative difficulty of individual attack categories changes across
models. \texttt{Ranking Manipulation}, for example, ranges from
86\% recall with GPT-5.5 to 17\% with Phi-4 and Mistral Nemo.
Similarly, \texttt{Metadata Gaps} ranges from 93\% with Gemini 3.1
Flash Lite and Claude Haiku~4.5 to 13\% with Mistral Nemo. Thus, no
single attack category is uniformly difficult across all judges.

\textbf{Answer to \ref{rq:synthetic}.}
Detection effectiveness varies substantially across discovery-time
attack classes and depends on the underlying semantic judge. Stronger
judges detect most attack categories with high recall, but individual
categories exhibit substantial model-dependent variation. These
results indicate that robustness to one class of Agent Card
manipulation does not necessarily imply robustness across the full
discovery-time attack surface.

\begin{table*}[!t]
\centering
  \begingroup
  \renewenvironment{table}[1][]{}{\par}%
  \renewenvironment{table*}[1][]{}{\par}%
  % Auto-generated by cross_model_eval_b.py — do not hand-edit.
\begin{table*}[t]
\centering
\caption{Cross-model security recall on the 260-card synthetic adversarial dataset.
  Security recall: MALICIOUS$\to$MALICIOUS; SUSPICIOUS$\to$\{SUSPICIOUS,MALICIOUS\}.
  There are no BENIGN cards, so this table admits no false-positive rate.
  ``—'' in the \emph{Params} row denotes undisclosed model size.}
\label{tab:cross_model_longform}
\resizebox{\textwidth}{!}{%
\renewcommand{\arraystretch}{1.15}
\setlength{\tabcolsep}{4pt}
\begin{tabular}{llr|rrrrrr|rrr}
\toprule
 & & & \multicolumn{6}{c}{\textit{Closed-Source}} & \multicolumn{3}{c}{\textit{Open-Source}} \\
\cmidrule(lr){4-9}
\cmidrule(lr){10-12}
\textbf{Category} & \textbf{Group} & \textbf{N} & \textbf{GPT-5.5} & \textbf{\shortstack{Claude\\Opus 4.8}} & \textbf{\shortstack{GPT-5.4\\mini}} & \textbf{\shortstack{Gemini 3.1\\Flash Lite}} & \textbf{\shortstack{Claude\\Haiku 4.5}} & \textbf{\shortstack{Gemini 2.5\\Flash Lite}} & \textbf{\shortstack{Llama\\3.1}} & \textbf{Phi-4} & \textbf{\shortstack{Mistral\\Nemo}} \\
\emph{Params} & & & — & — & — & — & — & — & 8B & 14B & 12B \\
\midrule
\texttt{indirect\_injection} & Semantic & 35 & 100\% & 97\% & 97\% & 100\% & 91\% & 89\% & 80\% & 54\% & 43\% \\
\texttt{ranking\_manipulation} & Semantic & 35 & 86\% & 57\% & 31\% & 69\% & 60\% & 57\% & 66\% & 17\% & 17\% \\
\texttt{skill\_deception} & Semantic & 35 & 89\% & 69\% & 89\% & 86\% & 77\% & 71\% & 57\% & 40\% & 23\% \\
\texttt{skill\_chaining} & Semantic & 35 & 100\% & 91\% & 91\% & 91\% & 89\% & 86\% & 66\% & 17\% & 20\% \\
\texttt{evaluator\_manipulation} & Semantic & 35 & 89\% & 74\% & 66\% & 77\% & 69\% & 63\% & 57\% & 54\% & 23\% \\
\addlinespace
\texttt{out\_of\_context\_skills} & Mixed & 25 & 96\% & 100\% & 92\% & 96\% & 100\% & 92\% & 60\% & 60\% & 28\% \\
\addlinespace
\texttt{url\_pivot} & Structural & 15 & 100\% & 100\% & 93\% & 100\% & 100\% & 87\% & 80\% & 73\% & 33\% \\
\texttt{unicode\_tricks} & Structural & 15 & 100\% & 93\% & 53\% & 80\% & 93\% & 53\% & 60\% & 47\% & 40\% \\
\texttt{context\_pollution} & Structural & 15 & 93\% & 67\% & 53\% & 80\% & 67\% & 40\% & 40\% & 33\% & 20\% \\
\texttt{metadata\_gaps} & Structural & 15 & 67\% & 80\% & 53\% & 93\% & 93\% & 40\% & 60\% & 60\% & 13\% \\
\midrule
\textbf{Overall recall} & & \textbf{260} & \textbf{92.3\%} & \textbf{81.5\%} & \textbf{73.8\%} & \textbf{86.5\%} & \textbf{81.9\%} & \textbf{70.8\%} & \textbf{63.5\%} & \textbf{42.7\%} & \textbf{25.8\%} \\
\quad MALICIOUS & & 175 & 92.6\% & 77.7\% & 74.9\% & 84.6\% & 77.1\% & 73.1\% & 65.1\% & 36.6\% & 25.1\% \\
\quad SUSPICIOUS & & 85 & 91.8\% & 89.4\% & 71.8\% & 90.6\% & 91.8\% & 65.9\% & 60.0\% & 55.3\% & 27.1\% \\
Accuracy (attacks) & & 260 & 77.3\% & 67.3\% & 57.7\% & 70.0\% & 66.5\% & 55.8\% & 45.4\% & 30.4\% & 19.6\% \\
\midrule
Avg.\ latency & & — & 4.5s & 4.0s & 1.3s & 1.3s & 2.8s & 1.1s & 8.9s & 16.9s & 8.4s \\
\bottomrule
\end{tabular}}
\end{table*}
  \endgroup
\end{table*}
% ------------------------------------------------------------------
\subsection{Cross-Model Evaluation (\ref{rq:models})}
\label{sec:evaluation:crossmodel}
% ------------------------------------------------------------------

To determine how strongly semantic attack detection depends on the
underlying LLM judge, we compare nine models on the same 260-card
Synthetic Adversarial Dataset. In addition to Claude Haiku~4.5, we
evaluate GPT-5.5, GPT-5.4-mini, Gemini~3.1 Flash Lite,
Gemini~2.5 Flash Lite, Claude Opus~4.8, Llama~3.1~8B, Phi-4~14B,
and Mistral~Nemo~12B. The open-weight models are run locally via
Ollama on an NVIDIA RTX~A2000 12~GB GPU. All models receive the
same semantic-validation prompt and are evaluated using the same
harness and severity-aware security-recall metric.
Table~\ref{tab:cross_model_longform} reports the results.

\noindent\textbf{Results.}
Semantic detection performance varies substantially across LLM
judges. GPT-5.5 achieves the highest overall security recall at
92.3\%, followed by Gemini~3.1 Flash Lite at 86.5\%, Claude
Haiku~4.5 at 81.9\%, and Claude Opus~4.8 at 81.5\%. GPT-5.4-mini
and Gemini~2.5 Flash Lite achieve 73.8\% and 70.8\%, respectively.
Among the evaluated open-weight models, Llama~3.1 reaches 63.5\%,
Phi-4 reaches 42.7\%, and Mistral~Nemo reaches 25.8\%. Overall,
security recall spans 66.5 percentage points across models evaluated
on the same cards and prompt.

Exact three-class attack accuracy is consistently lower than
severity-aware security recall. GPT-5.5, for example, achieves
92.3\% security recall but 77.3\% exact accuracy. 

The models also differ substantially in inference cost. Average
per-card latency ranges from 1.1 seconds for Gemini~2.5 Flash Lite
to 16.9 seconds for Phi-4 in our evaluation environment. Thus,
model selection affects not only detection effectiveness but also
the computational cost of pre-invocation validation.

\textbf{Answer to \ref{rq:models}.}
The choice of LLM judge affects semantic Agent Card
validation. Across the same 260 adversarial cards and evaluation
procedure, security recall ranges from 25.8\% to 92.3\%, while
inference latency also varies substantially. Consequently, the
semantic judge is an important deployment choice rather than an
interchangeable component of \framework.

\section{Discussion}
\label{sec:discussion}

This section places the proposed work in a broader context. We discuss its applicability to other agent-metadata protocols, examine the current limitations, and conclude by listing promising directions for future work.

% ------------------------------------------------------------------
\subsection{Generalizability}
\label{sec:discussion:generalization}

MCP~\cite{mcp2024spec} is the closest architectural analogue to A2A as both expose free-text metadata consumed by an orchestrator. Thus, our core finding generalizes: \emph{any agent-metadata protocol that relies on LLM interpretation of untrusted natural-language metadata inherits a similar attack surface.}

\noindent\textbf{Attack-category transfer.}
Eight of our ten categories transfer directly to MCP; \texttt{url\_pivot} maps partially (more naturally as tool-argument redirects~\cite{shi2026toolhijacker}) and \texttt{metadata\_gaps} must be adapted to MCP's schema. MCP can further increase impact via package-style distribution (supply-chain risk~\cite{ladisa2023sok}), local-privilege tools, and in-session tool-list updates that enable runtime injection.

\noindent\textbf{Architectural implication.}
The core \framework design is largely protocol-agnostic: structured
metadata can undergo protocol-specific conformance checks followed by
deterministic and semantic analysis before it influences autonomous
agent decisions. Applying this approach to other protocols would
require adapting structural validation to the corresponding metadata
schema and tailoring threat-detection features to protocol-specific
semantics. Cryptographic provenance mechanisms can complement this
analysis by authenticating the publisher and protecting metadata
integrity, but do not replace semantic validation of attacker-controlled
natural-language content.

% ------------------------------------------------------------------
\subsection{Limitations}
\label{sec:discussion:limits}

\begin{comment}

\noindent\textbf{Corpus scope.}
Our 310-card dataset is a census of observable A2A deployments at collection time (April 2026), not a sample from a mature ecosystem (7 valid cards per million probed domains). Corpus-level distributions should therefore be read as ``current state'' rather than long-run estimates.
    
\end{comment}

\textbf{Runtime Behavior.} Our framework focuses on securing the discovery-to-invocation process by evaluating Agent Cards prior to provider invocation. Our evaluation measures the detection of adversarial Agent Card content rather than end-to-end attack success against a deployed orchestrator.

While this design prevents many malicious or deceptive Agent Cards from influencing agent selection, it does not verify that a provider behaves consistently with its advertised capabilities after execution begins. A provider may
therefore present a benign Agent Card yet exhibit malicious or unexpected behavior during runtime. Integrating post-invocation behavioral monitoring with the proposed framework to continuously validate that observed behavior remains consistent with advertised capabilities is an important direction for future work.

\textbf{Authentication and Provenance.}
\framework evaluates the semantic trustworthiness of Agent Cards but does not replace cryptographic mechanisms for establishing their origin or integrity. The A2A specification defines an optional \texttt{signatures} field based on JSON Web Signatures (JWS), which enables providers to authenticate published Agent Cards. Such mechanisms are complementary to our approach: digital signatures establish \emph{who} published a card, whereas \framework evaluates \emph{what} the card communicates and whether its content is trustworthy for autonomous agent discovery. Consequently, a validly signed Agent Card may still contain deceptive or malicious metadata that requires semantic analysis.

\section{Future Work}
\label{sec:discussion:future}

This work represents an initial step toward securing the A2A discovery process through Agent Card validation. Future research should broaden the threat model to encompass coordinated multi-agent attacks, adaptive adversaries that modify Agent Cards over time, and attacks against the surrounding discovery infrastructure, including registries, caching services, and trust management mechanisms. Extending the framework to address these emerging threats would further strengthen the security of large-scale agent ecosystems.

\section{Conclusion}
\label{sec:conclusion}

The Agent2Agent (A2A) protocol places externally controlled
natural-language metadata on the critical path of autonomous agent
discovery and invocation. This work examined the resulting
pre-invocation trust problem through a taxonomy of ten discovery-time
Agent Card manipulation attacks and introduced \framework, a layered
validation framework for identifying potentially untrustworthy cards
before they influence orchestrator decisions.

Our evaluation combines an Internet-scale measurement of 310 deployed
Agent Cards with two controlled adversarial benchmarks: 660 cards
constructed by injecting attacks into real deployed cards and 260
synthetic adversarial cards spanning all ten attack classes. The
strongest semantic judge achieves 97.0\% security recall at a 1.7\%
false-positive rate on controlled attacks injected into real cards,
while cross-model security recall ranges from 25.8\% to 92.3\% on the
synthetic benchmark. These results demonstrate that discovery-time
Agent Card manipulation can be effectively identified before provider
invocation, while also showing that validation robustness depends
substantially on the underlying semantic judge.

As interoperable agent ecosystems expand, Agent Cards and similar
discovery metadata should be treated as a security boundary rather
than trusted service descriptions. Our results motivate
pre-invocation validation as a security primitive for autonomous agent
discovery and provide an empirical foundation for designing more
robust trust mechanisms for emerging multi-agent ecosystems.

\section*{Acknowledgment}
This work was supported in part by the National Science Foundation (NSF) under grant 2417062.

%\section{LLM Usage Statement}
%Large language Models are central to our research, but they were not used in any way to edit the present research document or produce any ideas; these are our own original. LLMs were used only to implement methods and algorithms, and all of them were author-validated for correctness.
% ================ MAIN BODY ENDS HERE ++==================================

% ============ APPENDIX SECTION ============
%\appendix
\counterwithin{figure}{section}
\counterwithin{table}{section}
\bibliographystyle{IEEEtran}
\bibliography{reference}

@misc{crewai_github_2026,
  author       = {{CrewAI, Inc.}},
  title        = {{CrewAI}: Framework for Orchestrating Role-Playing, Autonomous AI Agents},
  year         = {2026},
  howpublished = {\url{https://github.com/crewAIInc/crewAI}},
  note         = {GitHub repository. Accessed: 2026-07-15}
}

@online{langchain2026state,
  author       = {{LangChain}},
  title        = {State of Agent Engineering},
  year         = {2026},
  month        = jun,
  day          = {12},
  url          = {https://www.langchain.com/state-of-agent-engineering},
  note         = {Survey report. Accessed: 2026-07-15}
}

@article{dang2026multi,
  title={Multi-agent collaboration via evolving orchestration},
  author={Dang, Yufan and Qian, Chen and Luo, Xueheng and Fan, Jingru and Xie, Zihao and Shi, Ruijie and Chen, Weize and Yang, Cheng and Che, Xiaoyin and Tian, Ye and others},
  journal={Advances in neural information processing systems},
  volume={38},
  pages={165025--165059},
  year={2026}
}

@online{chopra2026orchestration,
  author       = {Nitasha Chopra},
  title        = {New and Improved: Multi-Agent Orchestration, Connected Experiences, and Faster Prompt Iteration},
  year         = {2026},
  month        = apr,
  day          = {1},
  organization = {Microsoft},
  url          = {https://www.microsoft.com/en-us/microsoft-copilot/blog/copilot-studio/new-and-improved-multi-agent-orchestration-connected-experiences-and-faster-prompt-iteration/},
  note         = {Microsoft Copilot Blog. Accessed: 2026-07-15}
}

@misc{commoncrawl2026march,
  author       = {{Common Crawl}},
  title        = {{Common Crawl March 2026 Crawl Archive (CC-MAIN-2026-12)}},
  year         = {2026},
  howpublished = {\url{https://data.commoncrawl.org/crawl-data/CC-MAIN-2026-12/index.html}},
  note         = {Accessed: 2026-09-23}
}

@article{louck,
title = "Improving Google A2A Protocol: Protecting Sensitive Data and Mitigating Unintended Harms in Multi-Agent Systems",
author = "Yedidel Louck and Ariel Stulman and Amit Dvir",
year = "2026",
month = jun,
day = "1",
doi = "10.1145/3821216",
journal = "ACM Transactions on Software Engineering and Methodology",
publisher = "Association for Computing Machinery",
}

@article{sharma2025towards,
  title={Towards unifying quantitative security benchmarking for multi agent systems},
  author={Sharma, Gauri and Kulkarni, Vidhi and King, Miles and Huang, Ken},
  journal={arXiv preprint arXiv:2507.21146},
  year={2025}
}

@INPROCEEDINGS{deshmukh2026agent,
  author={Deshmukh, Niharika and Rajurkar, Rani and Mamidwar, Piyush},
  booktitle={2026 International Conference on System, Computation, Automation and Networking (ICSCAN)}, 
  title={Agent Card Poisoning: Measuring Discovery-Layer Injection Attacks and Defenses in A2A Systems}, 
  year={2026},
  volume={},
  number={},
  pages={1-6},
  doi={10.1109/ICSCAN66520.2026.11588515}}

@inproceedings{li2026a2asecbench,
  title={A2asecbench: A protocol-aware security benchmark for agent-to-agent multi-agent systems},
  author={Li, Tianhao and Chu, Chuangxin and Zheng, Yujia and Zhang, Bohan and Gong, Neil and Xiao, Chaowei},
  booktitle={International Conference on Learning Representations},
  volume={2026},
  pages={122168--122191},
  year={2026}
}

@misc{perez2022ignore,
  author       = {Perez, F{\'a}bio and Ribeiro, Ian},
  title        = {Ignore Previous Prompt: Attack Techniques For Language Models},
  howpublished = {arXiv:2211.09527 / NeurIPS 2022 ML Safety Workshop},
  year         = {2022},
  month        = nov,
  url          = {https://arxiv.org/abs/2211.09527},
}

@article{he2024survey,
  author  = {He, Feng and Zhu, Tianqing and Ye, Dayong and Liu, Bo and Zhou, Wanlei and Yu, Philip S.},
  title   = {The Emerged Security and Privacy of {LLM} Agent: A Survey with Case Studies},
  journal = {ACM Computing Surveys},
  year    = {2025},
  doi     = {10.1145/3773080},
  url     = {https://arxiv.org/abs/2407.19354},
}

@inproceedings{greshake2023not,
  author    = {Greshake, Kai and Abdelnabi, Sahar and Mishra, Shailesh and Endres, Christoph and Holz, Thorsten and Fritz, Mario},
  title     = {Not What You've Signed Up For: Compromising Real-World {LLM}-Integrated Applications with Indirect Prompt Injection},
  booktitle = {Proceedings of the 16th ACM Workshop on Artificial Intelligence and Security ({AISec} '23)},
  publisher = {ACM},
  year      = {2023},
  doi       = {10.1145/3605764.3623985},
  url       = {https://arxiv.org/abs/2302.12173},
}

@inproceedings{liu2024formalizing,
  author    = {Liu, Yupei and Jia, Yuqi and Geng, Runpeng and Jia, Jinyuan and Gong, Neil Zhenqiang},
  title     = {Formalizing and Benchmarking Prompt Injection Attacks and Defenses},
  booktitle = {33rd USENIX Security Symposium (USENIX Security 24)},
  publisher = {USENIX Association},
  year      = {2024},
  url       = {https://www.usenix.org/conference/usenixsecurity24/presentation/liu-yupei},
}

@inproceedings{zhan2024injecagent,
  author    = {Zhan, Qiusi and Liang, Zhixiang and Ying, Zifan and Kang, Daniel},
  title     = {{InjecAgent}: Benchmarking Indirect Prompt Injections in Tool-Integrated Large Language Model Agents},
  booktitle = {Findings of the Association for Computational Linguistics: ACL 2024},
  publisher = {Association for Computational Linguistics},
  pages     = {10471--10506},
  year      = {2024},
  url       = {https://aclanthology.org/2024.findings-acl.624/},
}

@inproceedings{shi2026toolhijacker,
  author    = {Shi, Jiawen and Yuan, Zenghui and Tie, Guiyao and Zhou, Pan and Gong, Neil Zhenqiang and Sun, Lichao},
  title     = {Prompt Injection Attack to Tool Selection in {LLM} Agents},
  booktitle = {Proceedings of the Network and Distributed System Security Symposium (NDSS 2026)},
  year      = {2026},
  doi       = {10.14722/ndss.2026.s675},
  url       = {https://www.ndss-symposium.org/ndss-paper/prompt-injection-attack-to-tool-selection-in-llm-agents/},
}

@inproceedings{debenedetti2024agentdojo,
  author    = {Debenedetti, Edoardo and Zhang, Jie and Balunovi{\'c}, Mislav and Beurer-Kellner, Luca and Fischer, Marc and Tramèr, Florian},
  title     = {{AgentDojo}: A Dynamic Environment to Evaluate Prompt Injection Attacks and Defenses for {LLM} Agents},
  booktitle = {Advances in Neural Information Processing Systems 37 (NeurIPS 2024) Datasets and Benchmarks Track},
  year      = {2024},
  url       = {https://arxiv.org/abs/2406.13352},
}

@inproceedings{schick2023toolformer,
  author    = {Schick, Timo and Dwivedi-Yu, Jane and Dess{\`{\i}}, Roberto and Raileanu, Roberta and Lomeli, Maria and Zettlemoyer, Luke and Cancedda, Nicola and Scialom, Thomas},
  title     = {{Toolformer}: Language Models Can Teach Themselves to Use Tools},
  booktitle = {Advances in Neural Information Processing Systems 36 (NeurIPS 2023)},
  year      = {2023},
  url       = {https://arxiv.org/abs/2302.04761},
}

@inproceedings{yao2023react,
  author    = {Yao, Shunyu and Zhao, Jeffrey and Yu, Dian and Du, Nan and Shafran, Izhak and Narasimhan, Karthik and Cao, Yuan},
  title     = {{ReAct}: Synergizing Reasoning and Acting in Language Models},
  booktitle = {The Eleventh International Conference on Learning Representations (ICLR 2023)},
  year      = {2023},
  url       = {https://arxiv.org/abs/2210.03629},
}

@inproceedings{zheng2023judging,
  author    = {Zheng, Lianmin and Chiang, Wei-Lin and Sheng, Ying and Zhuang, Siyuan and Wu, Zhanghao and Zhuang, Yonghao and Lin, Zi and Li, Zhuohan and Li, Dacheng and Xing, Eric P. and Zhang, Hao and Gonzalez, Joseph E. and Stoica, Ion},
  title     = {Judging {LLM}-as-a-Judge with {MT-Bench} and {Chatbot Arena}},
  booktitle = {Advances in Neural Information Processing Systems 36 (NeurIPS 2023) Datasets and Benchmarks Track},
  year      = {2023},
  url       = {https://arxiv.org/abs/2306.05685},
}

@inproceedings{ladisa2023sok,
  author    = {Ladisa, Piergiorgio and Plate, Henrik and Martinez, Matias and Barais, Olivier},
  title     = {{SoK}: Taxonomy of Attacks on Open-Source Software Supply Chains},
  booktitle = {2023 IEEE Symposium on Security and Privacy (SP)},
  publisher = {IEEE},
  pages     = {1509--1526},
  year      = {2023},
  doi       = {10.1109/SP46215.2023.10179304},
  url       = {https://arxiv.org/abs/2204.04008},
}

@misc{nottingham2019rfc8615,
  author       = {Nottingham, Mark},
  title        = {Well-Known Uniform Resource Identifiers ({URIs})},
  howpublished = {RFC 8615, IETF},
  year         = {2019},
  month        = may,
  doi          = {10.17487/RFC8615},
  url          = {https://datatracker.ietf.org/doc/html/rfc8615},
}

@article{habler2025building,
  title={Building a secure agentic AI application leveraging A2A protocol},
  author={Habler, Idan and Huang, Ken and Narajala, Vineeth Sai and Kulkarni, Prashant},
  journal={arXiv preprint arXiv:2504.16902},
  year={2025}
}

@misc{a2a2026spec,
  author       = {{Agent2Agent Project} and {Linux Foundation}},
  title        = {Agent2Agent ({A2A}) Protocol Specification, version 1.0.0},
  howpublished = {Specification document, \url{https://a2a-protocol.org/latest/specification/}},
  year         = {2026},
  note         = {Authoritative Protobuf source: \url{https://github.com/a2aproject/A2A/blob/main/specification/a2a.proto}},
}

@misc{mcp2024spec,
  author       = {{Anthropic}},
  title        = {Model Context Protocol ({MCP}) Specification},
  howpublished = {Specification document, \url{https://modelcontextprotocol.io/specification}},
  year         = {2024},
}

@misc{a2aprotocolProtocol,
	author = {The Linux Foundation},
	title = {{A}2{A} {P}rotocol --- a2a-protocol.org},
	howpublished = {\url{https://a2a-protocol.org/latest/}},
	year = {},
	note = {[Accessed 15-07-2026]},
}

@misc{google2025a2a,
  author       = {{Google}},
  title        = {Agent2Agent ({A2A}): A New Era of Agent Interoperability},
  howpublished = {Google Developers Blog, \url{https://developers.googleblog.com/en/a2a-a-new-era-of-agent-interoperability/}},
  year         = {2025},
  month        = apr,
  note         = {Protocol subsequently contributed to the Linux Foundation, June 2025.
                  Source repository: \url{https://github.com/a2aproject/A2A}},
}

@online{langchain2026,
  author  = {{LangChain}},
  title   = {LangChain},
  year    = {2026},
  url     = {https://www.langchain.com/},
  urldate = {2026-07-16}
}

\end{document}